\documentclass[twocolumn,letterpaper,aps,prc,longbibliography,superscriptaddress,nofootinbib,floatfix]{revtex4-2}

\usepackage{graphicx}	
\usepackage{multirow}
\usepackage{xspace}	

\newcommand{\zbbc}{\mbox{$z_{\rm BBC}$}\xspace}
\newcommand{\pt}{\mbox{$p_T$}\xspace}
\newcommand{\pz}{\mbox{$p_z$}\xspace}

\newcommand{\rpa}{\mbox{$R_{pA}$}\xspace}
\newcommand{\npart}{\mbox{$\langle N_{\rm part} \rangle$}\xspace}
\newcommand{\ncoll}{\mbox{$\langle N_{\rm coll} \rangle$}\xspace}

\newcommand{\sqs}{\mbox{$\sqrt{s}$}\xspace}
\newcommand{\sqstwo}{\mbox{$\sqrt{s}=200$~GeV}\xspace}
\newcommand{\sqsn}{\mbox{$\sqrt{s_{_{NN}}}$}\xspace}
\newcommand{\sqsntwo}{\mbox{$\sqrt{s_{_{NN}}}=200$~GeV}\xspace}
\newcommand{\pp}{\mbox{$p$+$p$}\xspace}
\newcommand{\dau}{\mbox{$d$$+$Au}\xspace}

\newcommand{\pau}{\mbox{$p$$+$Au}\xspace}
\newcommand{\pal}{\mbox{$p$$+$Al}\xspace}
\newcommand{\palau}{\mbox{$p$$+$Al and $p$$+$Au}\xspace}
\newcommand{\palaup}{\mbox{$p$$+$Al, $p$$+$Au, and $p$+$p$}\xspace}
\newcommand{\auau}{\mbox{Au$+$Au}\xspace}

\newcommand{\geant}{\mbox{{\sc geant4}}\xspace}
\newcommand{\pythia}{\mbox{{\sc pythia}}\xspace}
\newcommand{\hijing}{\mbox{{\sc hijing}}\xspace}
\newcommand{\cteq}{\mbox{{\sc cteq6l}}\xspace}
\newcommand{\ncteq}{\mbox{n{\sc cteq15}}\xspace}
\newcommand{\epps}{\mbox{{\sc epps16}}\xspace}
\newcommand{\xbj}{\mbox{$x_{\rm Bj}$}\xspace}
\newcommand{\minitab}[2][l]{\begin{tabular}{#1}#2\end{tabular}}

\begin{document}

\title{Nuclear-modification factor of charged hadrons at forward and 
backward rapidity in $p$$+$Al and $p$$+$Au collisions at 
$\sqrt{s_{_{NN}}}=200$\,GeV}

\newcommand{\abilene}{Abilene Christian University, Abilene, Texas 79699, USA}
\newcommand{\augie}{Department of Physics, Augustana University, Sioux Falls, South Dakota 57197, USA}
\newcommand{\banaras}{Department of Physics, Banaras Hindu University, Varanasi 221005, India}
\newcommand{\barc}{Bhabha Atomic Research Centre, Bombay 400 085, India}
\newcommand{\baruch}{Baruch College, City University of New York, New York, New York, 10010 USA}
\newcommand{\bnlcoll}{Collider-Accelerator Department, Brookhaven National Laboratory, Upton, New York 11973-5000, USA}
\newcommand{\bnlphys}{Physics Department, Brookhaven National Laboratory, Upton, New York 11973-5000, USA}
\newcommand{\caucr}{University of California-Riverside, Riverside, California 92521, USA}
\newcommand{\charlesczech}{Charles University, Ovocn\'{y} trh 5, Praha 1, 116 36, Prague, Czech Republic}
\newcommand{\cns}{Center for Nuclear Study, Graduate School of Science, University of Tokyo, 7-3-1 Hongo, Bunkyo, Tokyo 113-0033, Japan}
\newcommand{\colorado}{University of Colorado, Boulder, Colorado 80309, USA}
\newcommand{\columbia}{Columbia University, New York, New York 10027 and Nevis Laboratories, Irvington, New York 10533, USA}
\newcommand{\czechtech}{Czech Technical University, Zikova 4, 166 36 Prague 6, Czech Republic}
\newcommand{\debrecen}{Debrecen University, H-4010 Debrecen, Egyetem t{\'e}r 1, Hungary}
\newcommand{\elte}{ELTE, E{\"o}tv{\"o}s Lor{\'a}nd University, H-1117 Budapest, P{\'a}zm{\'a}ny P.~s.~1/A, Hungary}
\newcommand{\eszterhazy}{Eszterh\'azy K\'aroly University, K\'aroly R\'obert Campus, H-3200 Gy\"ongy\"os, M\'atrai \'ut 36, Hungary}
\newcommand{\ewha}{Ewha Womans University, Seoul 120-750, Korea}
\newcommand{\famu}{Florida A\&M University, Tallahassee, FL 32307, USA}
\newcommand{\fsu}{Florida State University, Tallahassee, Florida 32306, USA}
\newcommand{\gsu}{Georgia State University, Atlanta, Georgia 30303, USA}
\newcommand{\hiroshima}{Hiroshima University, Kagamiyama, Higashi-Hiroshima 739-8526, Japan}
\newcommand{\howard}{Department of Physics and Astronomy, Howard University, Washington, DC 20059, USA}
\newcommand{\ihepprot}{IHEP Protvino, State Research Center of Russian Federation, Institute for High Energy Physics, Protvino, 142281, Russia}
\newcommand{\illuiuc}{University of Illinois at Urbana-Champaign, Urbana, Illinois 61801, USA}
\newcommand{\inrras}{Institute for Nuclear Research of the Russian Academy of Sciences, prospekt 60-letiya Oktyabrya 7a, Moscow 117312, Russia}
\newcommand{\instpasczech}{Institute of Physics, Academy of Sciences of the Czech Republic, Na Slovance 2, 182 21 Prague 8, Czech Republic}
\newcommand{\isu}{Iowa State University, Ames, Iowa 50011, USA}
\newcommand{\jaea}{Advanced Science Research Center, Japan Atomic Energy Agency, 2-4 Shirakata Shirane, Tokai-mura, Naka-gun, Ibaraki-ken 319-1195, Japan}
\newcommand{\jeonbuk}{Jeonbuk National University, Jeonju, 54896, Korea}
\newcommand{\kek}{KEK, High Energy Accelerator Research Organization, Tsukuba, Ibaraki 305-0801, Japan}
\newcommand{\korea}{Korea University, Seoul, 02841}
\newcommand{\kurchatov}{National Research Center ``Kurchatov Institute", Moscow, 123098 Russia}
\newcommand{\kyoto}{Kyoto University, Kyoto 606-8502, Japan}
\newcommand{\lawllnl}{Lawrence Livermore National Laboratory, Livermore, California 94550, USA}
\newcommand{\losalamos}{Los Alamos National Laboratory, Los Alamos, New Mexico 87545, USA}
\newcommand{\lund}{Department of Physics, Lund University, Box 118, SE-221 00 Lund, Sweden}
\newcommand{\lyon}{IPNL, CNRS/IN2P3, Univ Lyon, Université Lyon 1, F-69622, Villeurbanne, France}
\newcommand{\maryland}{University of Maryland, College Park, Maryland 20742, USA}
\newcommand{\mass}{Department of Physics, University of Massachusetts, Amherst, Massachusetts 01003-9337, USA}
\newcommand{\michigan}{Department of Physics, University of Michigan, Ann Arbor, Michigan 48109-1040, USA}
\newcommand{\muhlenberg}{Muhlenberg College, Allentown, Pennsylvania 18104-5586, USA}
\newcommand{\nara}{Nara Women's University, Kita-uoya Nishi-machi Nara 630-8506, Japan}
\newcommand{\natmephi}{National Research Nuclear University, MEPhI, Moscow Engineering Physics Institute, Moscow, 115409, Russia}
\newcommand{\newmex}{University of New Mexico, Albuquerque, New Mexico 87131, USA}
\newcommand{\nmsu}{New Mexico State University, Las Cruces, New Mexico 88003, USA}
\newcommand{\northcg}{Physics and Astronomy Department, University of North Carolina at Greensboro, Greensboro, North Carolina 27412, USA}
\newcommand{\ohio}{Department of Physics and Astronomy, Ohio University, Athens, Ohio 45701, USA}
\newcommand{\ornl}{Oak Ridge National Laboratory, Oak Ridge, Tennessee 37831, USA}
\newcommand{\orsay}{IPN-Orsay, Univ.~Paris-Sud, CNRS/IN2P3, Universit\'e Paris-Saclay, BP1, F-91406, Orsay, France}
\newcommand{\peking}{Peking University, Beijing 100871, People's Republic of China}
\newcommand{\pnpi}{PNPI, Petersburg Nuclear Physics Institute, Gatchina, Leningrad region, 188300, Russia}
\newcommand{\riken}{RIKEN Nishina Center for Accelerator-Based Science, Wako, Saitama 351-0198, Japan}
\newcommand{\rikjrbrc}{RIKEN BNL Research Center, Brookhaven National Laboratory, Upton, New York 11973-5000, USA}
\newcommand{\rikkyo}{Physics Department, Rikkyo University, 3-34-1 Nishi-Ikebukuro, Toshima, Tokyo 171-8501, Japan}
\newcommand{\saispbstu}{Saint Petersburg State Polytechnic University, St.~Petersburg, 195251 Russia}
\newcommand{\seoulnat}{Department of Physics and Astronomy, Seoul National University, Seoul 151-742, Korea}
\newcommand{\stonybrkc}{Chemistry Department, Stony Brook University, SUNY, Stony Brook, New York 11794-3400, USA}
\newcommand{\stonycrkp}{Department of Physics and Astronomy, Stony Brook University, SUNY, Stony Brook, New York 11794-3800, USA}
\newcommand{\tenn}{University of Tennessee, Knoxville, Tennessee 37996, USA}
\newcommand{\titech}{Department of Physics, Tokyo Institute of Technology, Oh-okayama, Meguro, Tokyo 152-8551, Japan}
\newcommand{\tsukuba}{Tomonaga Center for the History of the Universe, University of Tsukuba, Tsukuba, Ibaraki 305, Japan}
\newcommand{\vandy}{Vanderbilt University, Nashville, Tennessee 37235, USA}
\newcommand{\weizmann}{Weizmann Institute, Rehovot 76100, Israel}
\newcommand{\wigner}{Institute for Particle and Nuclear Physics, Wigner Research Centre for Physics, Hungarian Academy of Sciences (Wigner RCP, RMKI) H-1525 Budapest 114, POBox 49, Budapest, Hungary}
\newcommand{\yonsei}{Yonsei University, IPAP, Seoul 120-749, Korea}
\newcommand{\zagreb}{Department of Physics, Faculty of Science, University of Zagreb, Bijeni\v{c}ka c.~32 HR-10002 Zagreb, Croatia}
\affiliation{\abilene}
\affiliation{\augie}
\affiliation{\banaras}
\affiliation{\barc}
\affiliation{\baruch}
\affiliation{\bnlcoll}
\affiliation{\bnlphys}
\affiliation{\caucr}
\affiliation{\charlesczech}
\affiliation{\cns}
\affiliation{\colorado}
\affiliation{\columbia}
\affiliation{\czechtech}
\affiliation{\debrecen}
\affiliation{\elte}
\affiliation{\eszterhazy}
\affiliation{\ewha}
\affiliation{\famu}
\affiliation{\fsu}
\affiliation{\gsu}
\affiliation{\hiroshima}
\affiliation{\howard}
\affiliation{\ihepprot}
\affiliation{\illuiuc}
\affiliation{\inrras}
\affiliation{\instpasczech}
\affiliation{\isu}
\affiliation{\jaea}
\affiliation{\jeonbuk}
\affiliation{\kek}
\affiliation{\korea}
\affiliation{\kurchatov}
\affiliation{\kyoto}
\affiliation{\lawllnl}
\affiliation{\losalamos}
\affiliation{\lund}
\affiliation{\lyon}
\affiliation{\maryland}
\affiliation{\mass}
\affiliation{\michigan}
\affiliation{\muhlenberg}
\affiliation{\nara}
\affiliation{\natmephi}
\affiliation{\newmex}
\affiliation{\nmsu}
\affiliation{\northcg}
\affiliation{\ohio}
\affiliation{\ornl}
\affiliation{\orsay}
\affiliation{\peking}
\affiliation{\pnpi}
\affiliation{\riken}
\affiliation{\rikjrbrc}
\affiliation{\rikkyo}
\affiliation{\saispbstu}
\affiliation{\seoulnat}
\affiliation{\stonybrkc}
\affiliation{\stonycrkp}
\affiliation{\tenn}
\affiliation{\titech}
\affiliation{\tsukuba}
\affiliation{\vandy}
\affiliation{\weizmann}
\affiliation{\wigner}
\affiliation{\yonsei}
\affiliation{\zagreb}
\author{C.~Aidala} \affiliation{\michigan} 
\author{Y.~Akiba} \email[PHENIX Spokesperson: ]{akiba@rcf.rhic.bnl.gov} \affiliation{\riken} \affiliation{\rikjrbrc} 
\author{M.~Alfred} \affiliation{\howard} 
\author{V.~Andrieux} \affiliation{\michigan} 
\author{N.~Apadula} \affiliation{\isu} 
\author{H.~Asano} \affiliation{\kyoto} \affiliation{\riken} 
\author{B.~Azmoun} \affiliation{\bnlphys} 
\author{V.~Babintsev} \affiliation{\ihepprot} 
\author{N.S.~Bandara} \affiliation{\mass} 
\author{K.N.~Barish} \affiliation{\caucr} 
% ...
\collaboration{PHENIX Collaboration} \noaffiliation

\date{\today}


\begin{abstract}


The PHENIX experiment has studied nuclear effects in $p$$+$Al and 
$p$$+$Au collisions at $\sqrt{s_{_{NN}}}=200$\,GeV on charged hadron 
production at forward rapidity ($1.4<\eta<2.4$, $p$-going direction) and 
backward rapidity ($-2.2<\eta<-1.2$, $A$-going direction).  Such effects 
are quantified by measuring nuclear modification factors as a function 
of transverse momentum and pseudorapidity in various collision 
multiplicity selections. In central $p$$+$Al and $p$$+$Au collisions, a 
suppression (enhancement) is observed at forward (backward) rapidity 
compared to the binary scaled yields in $p$+$p$ collisions. The 
magnitude of enhancement at backward rapidity is larger in $p$$+$Au 
collisions than in $p$$+$Al collisions, which have a smaller number of 
participating nucleons. However, the results at forward rapidity show a 
similar suppression within uncertainties. The results in the integrated 
centrality are compared with calculations using nuclear parton 
distribution functions, which show a reasonable agreement at the forward 
rapidity but fail to describe the backward rapidity enhancement.

\end{abstract}

\maketitle

\section{Introduction}
\label{sec:introduction}

Measurements of particle production in heavy-ion collisions enable the 
study of properties of a hot and dense nuclear medium called the 
quark-gluon plasma 
(QGP)~\cite{Adcox:2004mh,Adams:2005dq,Back:2004je,Arsene:2004fa}. An initial 
striking observation at the Relativistic Heavy Ion Collider (RHIC) was 
that production of high transverse momentum (\pt) hadrons in \auau 
collisions is strongly suppressed compared to that in \pp collisions 
scaled by the number of binary collisions.  This suppression indicates 
that partons experience substantial energy loss as they traverse the 
QGP, a phenomenon called jet-quenching~\cite{Gyulassy:2003mc}. A control 
experiment involving a deuteron projectile on a heavy-ion target, \dau, 
was carried out to test whether the feature of strong energy loss is 
still present in a collision system of much smaller size. The results in 
\dau collisions at midrapidity presented in Ref.~\cite{Adler:2003ii} 
showed no suppression at high \pt, initially leading to the conclusion 
that QGP itself---and associated jet quenching---were unique to 
collisions of larger heavy ions. In the ten years because these initial 
measurements, indications of QGP formation in smaller collision systems 
including \dau have been found, though without evidence of jet quenching 
phenomena~\cite{Nagle:2018nvi}.

Although there were no indications of strong suppression of high \pt 
particles in \dau collisions, detailed measurements do indicate other 
particle-production modifications relative to \pp 
collisions~\cite{Arsene:2004ux,Adams:2006uz,Adare:2011sc,Adare:2013esx,Adare:2013lkk}. 
At midrapidity, a centrality-dependent enhancement of charged hadron 
production was observed at intermediate \pt ($2<\pt<5~{\rm 
GeV}/c$)~\cite{Adare:2013esx} in \dau collisions at \sqsntwo.  These 
nuclear effects may be due to initial- and/or final-state multiple 
scatterings of incoming and outgoing 
partons~\cite{Accardi:2002ik,Kang:2014hha}. Processes such as radial 
flow~\cite{Hirano:2003pw} and recombination~\cite{Hwa:2002tu} developed 
for heavy-ion collisions were also investigated to explain a stronger 
enhancement of $p$ and $\bar{p}$ over $\pi^{\pm}$ and 
$K^{\pm}$~\cite{Adare:2013esx}. Recent results of collectivity amongst 
identified particles in small collision systems at RHIC and the Large 
Hadron Collider~\cite{Nagle:2018nvi} have been also explained 
within the hydrodynamic evolution 
model~\cite{Habich:2014jna,Shen:2016zpp}.

The study of particle production at forward and backward rapidity can 
provide additional information on nuclear effects such as initial-state 
energy loss~\cite{Qiu:2004da} and modification of nuclear parton 
distribution functions 
(nPDF)~\cite{Geesaman:1995yd,deFlorian:2011fp,Kovarik:2015cma,Eskola:2016oht,AbdulKhalek:2019mzd}. 
Of particular interest are gluons at small Bjorken \xbj (fraction of the 
proton's longitudinal momentum carried by the parton), where the 
dramatic increase of gluon density leads to expectation of saturation. 
This is often described within the color glass condensate (CGC) 
framework~\cite{McLerran:1993ka}. A strong centrality dependent 
suppression of single and dihadron production has been observed at 
forward rapidity in \dau collisions at 
\sqsntwo~\cite{Arsene:2004ux,Adams:2006uz,Adare:2011sc}. A CGC 
calculation provides a good description of the experimental 
data~\cite{Marquet:2007vb,Albacete:2010bs}. Also, a perturbative quantum 
chromodynamics (pQCD) calculation considering coherent multiple 
scattering with small-\xbj gluons reproduces the suppression of 
particle production at forward rapidity~\cite{Qiu:2004da,Kang:2011bp}. 
Another very different explanation for the suppression at forward 
rapidity is that color fluctuation effects modify the size of the 
high-\xbj partons in the proton~\cite{Alvioli:2014eda,Alvioli:2017wou}.

Accessible quark and gluon \xbj ranges depend on the pseudorapidity 
($\eta$) and transverse momentum of final state hadrons or jets. 
Therefore, measurements over a wide kinematic range are quite useful to 
further understand nuclear effects in small collision systems. PHENIX 
experiment has two muon spectrometers that provide wide coverage at 
forward ($\xbj \approx 0.02$, shadowing region) and backward rapidity 
($\xbj \approx 0.1$, anti-shadowing region). In the previous study of 
nuclear effects on charged hadron production in \dau collisions at 
\sqsntwo~\cite{Adler:2004eh}, a significant suppression was observed at 
forward rapidity in high multiplicity collisions compared to that in low 
multiplicity collisions, whereas a moderate enhancement is seen at 
backward rapidity. Although the direction of modification is consistent 
with the expectation from nPDF modification, no specific model 
comparison was presented.

High statistics data samples of \pp, \pal, and \pau collisions at 
\sqsntwo were collected in 2015 by PHENIX. These data samples combined 
with the availability of a new forward silicon vertex tracking 
detectors, which enable the selection of particle tracks coming from the 
collision point, significantly improved \pt and $\eta$ resolutions. The 
charged hadron analysis with these data sets can extend the previous 
study in \dau collisions~\cite{Adler:2004eh}, and a comparison between 
\pal and \pau of very different size of nuclei can provide new 
information on nuclear effects on charged hadron production in $p$$+$$A$ 
collisions.

In this paper, we present nuclear modification factors of charged hadron 
production at forward and backward rapidity in \pal and \pau collisions 
at \sqsntwo of various multiplicities.  Section~\ref{sec:experiment} 
describes the experimental setup and the data sets used in this 
analysis.  Section~\ref{sec:analysis} details the analysis methods.  
Section~\ref{sec:sys_uncert} discusses systematic uncertainties.
Section~\ref{sec:results} presents results and discussion.
Section~\ref{sec:summary} gives the summary and conclusions.

\section{Experimental Setup}
\label{sec:experiment}

\begin{figure}[thb]
\includegraphics[width=0.98\linewidth]{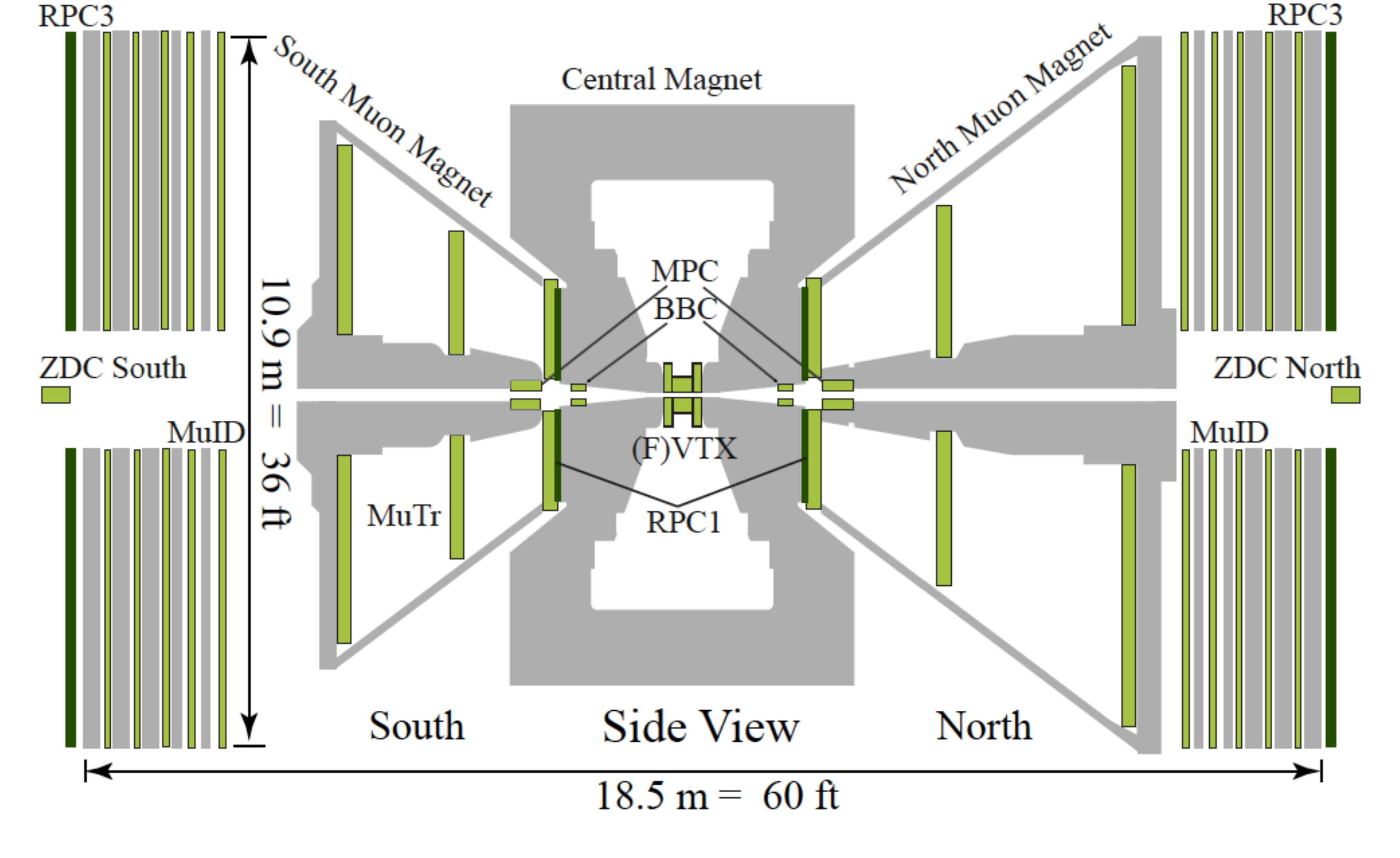}
\caption{\label{fig:PHENIX}
Side view of the PHENIX detector in 2015.}
\end{figure}

The PHENIX detector~\cite{Adcox:2003zm} comprises two central arm 
spectrometers at midrapidity and two muon arm spectrometers at forward 
and backward rapidity.  The detector configuration during the data 
taking in 2015 is shown in Fig.~\ref{fig:PHENIX}. The muon spectrometers 
have full-azimuthal acceptance, covering $-2.2<\eta<-1.2$ (south arm) 
and $1.2<\eta<2.4$ (north arm). Each muon arm comprises a forward 
silicon vertex tracker (FVTX), followed by a hadron absorber and a muon 
spectrometer. The muon spectrometer is composed of a muon tracker (MuTr) 
embedded in a magnetic field followed by a muon identifier (MuID).

The FVTX is a silicon detector with four stations in each arm.  Each 
station comprises 96 sensors along the $\phi$ direction. Each silicon 
sensor is finely segmented along the radial direction, with a strip 
pitch of $75~\mu{\rm m}$. The primary purpose of the FVTX is to measure 
a precise collision vertex also constrained by the silicon vertex 
tracker (VTX) at midrapidity. The FVTX was also designed to measure 
precise momentum vector information of charged particles entering the 
muon spectrometer before suffering large multiple scattering in the 
hadron absorber. More technical details on the FVTX are available in 
Ref.~\cite{Aidala:2013vna}. Following the FVTX is the hadron absorber, 
composed of layers of copper, iron, and stainless steel, corresponding 
to 7.2 nuclear interaction lengths ($\lambda_{I}$).  Hadrons entering 
the absorber are suppressed by a factor of approximately 1000, thus 
significantly reducing hadronic background for muon-based measurements.

The MuTr has two arms each consisting of three stations of cathode strip 
chambers, which are inside a magnet with a radial field integral of 
$\int{{\rm B}\cdot dl}=0.72~T\cdot{\rm m}$.  The MuTr provides a 
momentum measurement for charged particles. The MuID is composed of five 
layers (referred to as gap 0--4) of steel absorber (4.8 (5.4) 
$\lambda_{I}$ for south (north) arm) and two planes of Iarocci tubes.  
This enables the separation of muons and hadrons based on their 
penetration depth at a given reconstructed momentum. The MuTr and MuID 
are also used to trigger events containing at least one muon or hadron 
candidate. The MuID trigger is designed to enrich events with muons by 
requiring at least one hit in either gap 3 or 4. Hadrons that stop only 
after partially penetrating the MuID can be enhanced by requiring no hit 
in gap 4. The MuTr trigger is used to sample high momentum tracks by 
requiring a track sagitta less than three MuTr cathode strips wide at 
the middle station of the MuTr.  A more detailed discussion of the 
PHENIX muon arms can be found in 
Ref.~\cite{Akikawa:2003zs,Adachi:2013qha}.

The beam-beam counters (BBC)~\cite{Allen:2003zt} comprise two arrays 
of 64 quartz \v{C}erenkov detectors located at $z=\pm144~{\rm cm}$ from 
the nominal interaction point. Each BBC has an acceptance covering the 
full azimuth and $3.1<|\eta|<3.9$. The BBCs are used to determine the 
collision-vertex position along the beam axis ($z_{\rm BBC}$) with a 
resolution of roughly 2~cm in \pp collisions. They also provide a 
minimum bias (MB) trigger by requiring at least one hit in each BBC. The 
BBC trigger efficiency, determined from the Van der Meer scan 
technique~\cite{Drees:2003zza}, is 55\% for inelastic \pp events and 
79\% for events with midrapidity particle 
production~\cite{Adler:2003pb}. In \palau collisions, charged particle 
multiplicity in BBC in the Al- and Au-going direction ($-4.9<\eta<-3.1$) is 
used to categorize the event centrality. The BBC trigger is for 72\% 
(84\%) of inelastic \pal (\pau) collisions. Centrality dependent bias 
factors to account for the efficiency for MB triggered events and hard 
scattering events have been obtained based on the method developed in 
Ref.~\cite{Adare:2013nff}.

\section{Data analysis}
\label{sec:analysis}

\subsection{Data set}
\label{sec:dataset}

Data sets used in this analysis include \pp, \pal, and \pau collisions 
at \sqsntwo collected with the PHENIX detector in 2015. Events are 
required to have $|\zbbc|<20$~cm.  The improved precision vertex from 
the silicon trackers (VTX and FVTX) is not used in this analysis due to 
the track multiplicity-dependent vertex reconstruction efficiency. The 
analyzed event samples are required to have at least one track candidate 
in the MuTr and MuID satisfying either single hadron or single muon 
trigger in coincidence with the MB trigger. The integrated luminosity of 
the data used in this analysis is $23~{\rm pb}^{-1}$ in \pp, 260 ${\rm 
nb^{-1}}$ in \pal, and 80 ${\rm nb^{-1}}$ in \pau collisions.

\subsection{Hadron selection}
\label{sec:hadronsel}

\begin{figure*}[thb]
\includegraphics[width=0.998\linewidth]{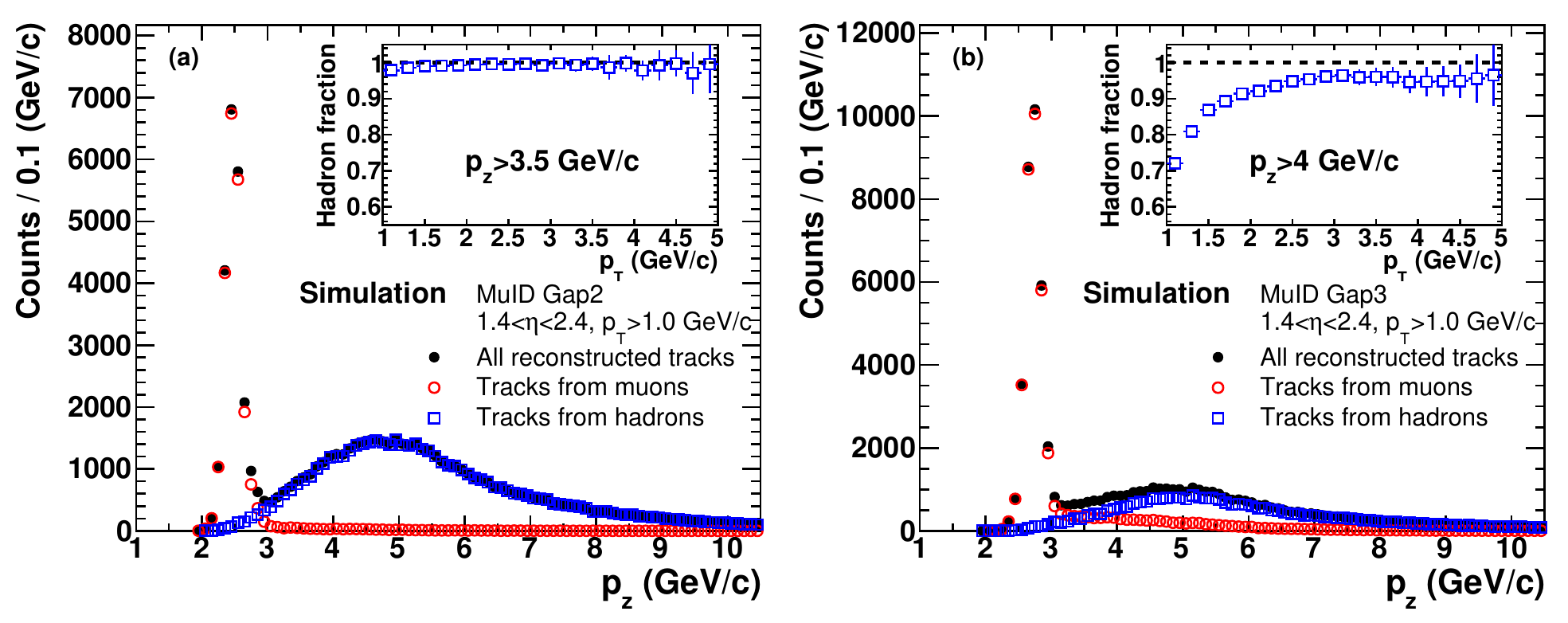}
\caption{\label{fig:g4sim_pz}
\geant-detector-simulation results for the \pz distributions of 
reconstructed tracks at the MuID (a) gap 2 and (b) gap 3 in the 
north muon arm.  The insets show the fraction of hadrons as a 
function of \pt with a \pz cut to help reject muon track candidates.}
\end{figure*}

The majority of hadrons emitted from the collision are stopped inside 
the hadron absorber.  Hadrons which pass through the hadron absorber 
enter the MuTr and can still be stopped in the middle of the MuID by 
producing hadronic showers in the additional steel absorber planes. Low 
momentum muons can also be stopped due to ionization energy loss, but 
the momentum distribution measured in the MuTr is very different for 
these muons and hadrons which are stopped in the MuID. 
Figure~\ref{fig:g4sim_pz} shows the longitudinal momentum (\pz) 
distributions of reconstructed tracks at the north arm MuID gaps 2 and 3 
from a full \geant detector simulation of charged hadrons (see 
Sec.~\ref{sec:acceff}). Muon tracks from light hadron decays show a 
narrow \pz distribution in $2.5<\pz<3.0~{\rm GeV}/c$, whereas tracks 
from hadrons show a much broader distribution. Therefore, tracks from 
hadrons can be enriched with a proper \pz cut ($3.5~{\rm GeV}/c$ for gap 
2 and $4~{\rm GeV}/c$ for gap 3).  The inset plots show the hadron 
fraction as a function of \pt with the \pz cuts. The hadron purity is 
$>98\%$ ($>90\%$) at MuID gap 2 (gap 3) for $\pt>1.5~{\rm GeV}/c$. The 
contamination of muons in the combined sample for both MuID gap 2 and 
gap 3 is less than 5\% based on this simulation study.

\begin{figure*}[thb]
\includegraphics[width=0.998\linewidth]{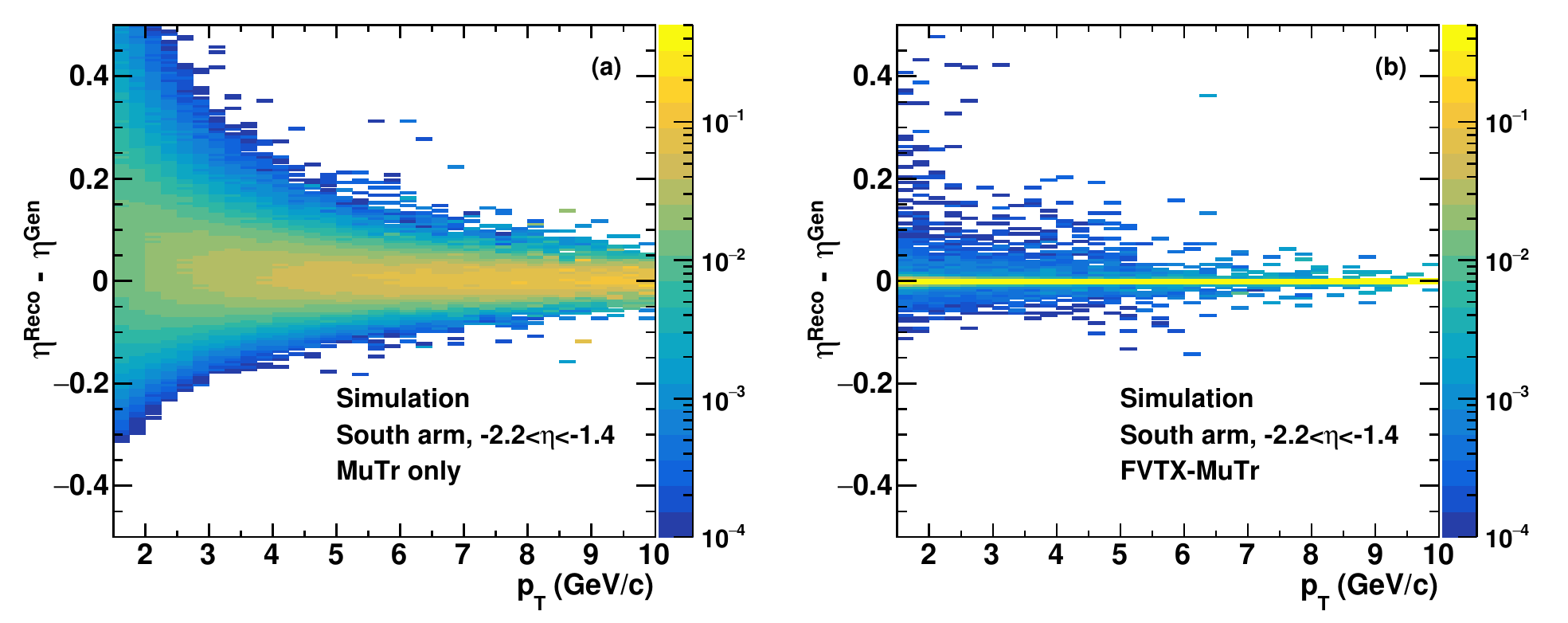}
\caption{\label{fig:g4sim_deta}
Comparison of the \pt-dependent $\eta$ resolution of tracks at MuID 
gap 2 and gap 3 in the south arm between tracks reconstructed with 
(a) MuTr only and (b) FVTX-MuTr association.}
\end{figure*}

One benefit from the FVTX is that the initial momentum vector of hadrons 
can be measured precisely before they undergo significant multiple 
scattering inside the absorber. In particular, the FVTX has very fine 
segmentation in the radial direction which can improve the \pt and 
$\eta$ resolution of measured tracks, both of which are important for 
this analysis. Figure~\ref{fig:g4sim_deta} shows the $\Delta\eta$ 
distribution between reconstructed tracks ($\eta^{\rm Reco}$) and 
true tracks ($\eta^{\rm Gen}$) as a function of \pt for hadron 
candidates from the \geant simulation. In the case where momentum 
information from only the MuTr is used, shown in 
Fig.~\ref{fig:g4sim_deta} (a), the smearing in $\eta$ is quite large.  
This is significantly improved by requiring association with FVTX 
tracks, shown in Fig.~\ref{fig:g4sim_deta} (b).

\subsection{Trigger efficiency}
\label{sec:trigeff}

\begin{figure}[htb]
\includegraphics[width=1.00\linewidth]{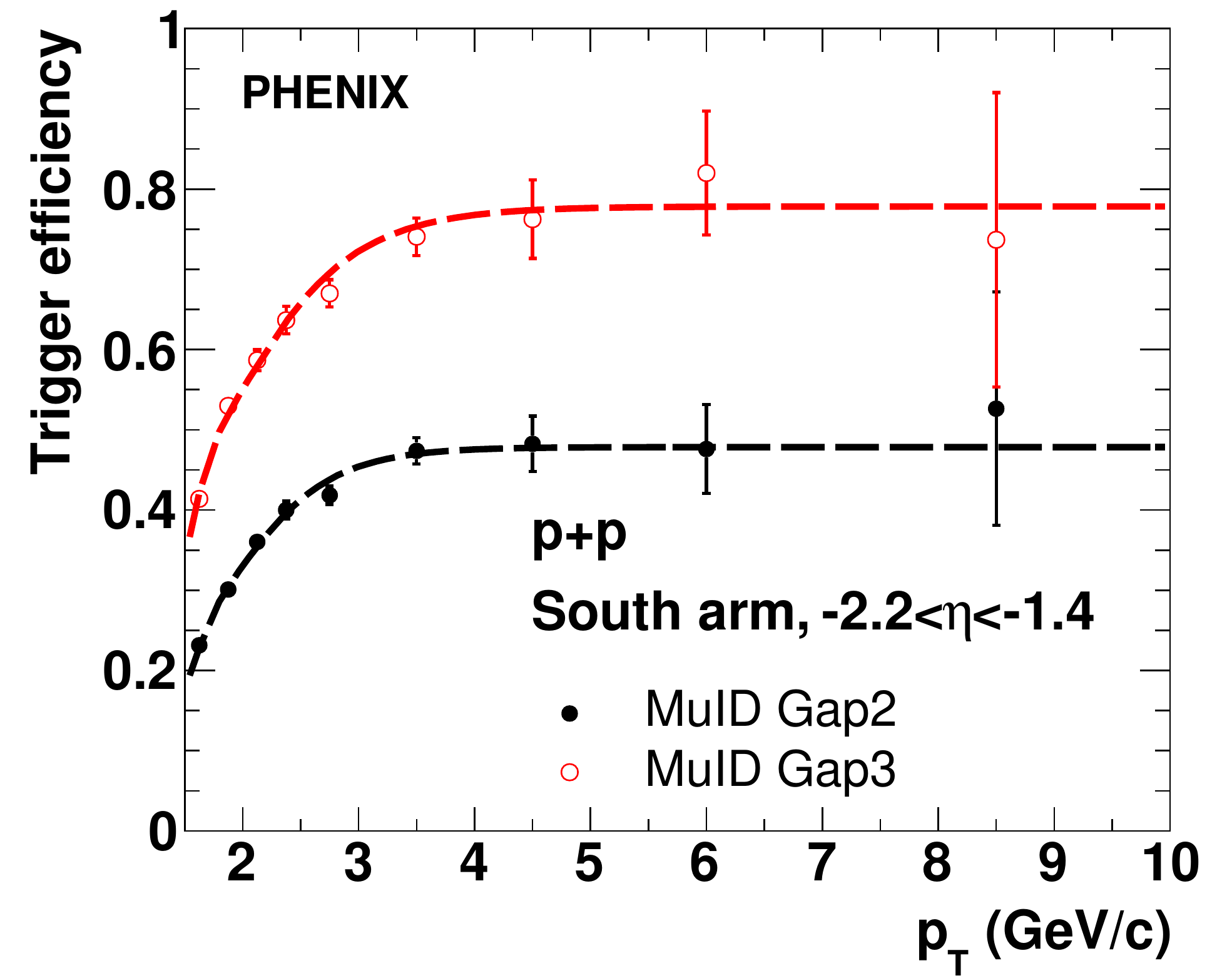}
\caption{\label{fig:trigeff_pp}
Trigger efficiency of hadron candidates as a function of \pt in the 
south arm evaluated in \pp collisions.}
\end{figure}

\begin{figure}[htb]
\includegraphics[width=1.00\linewidth]{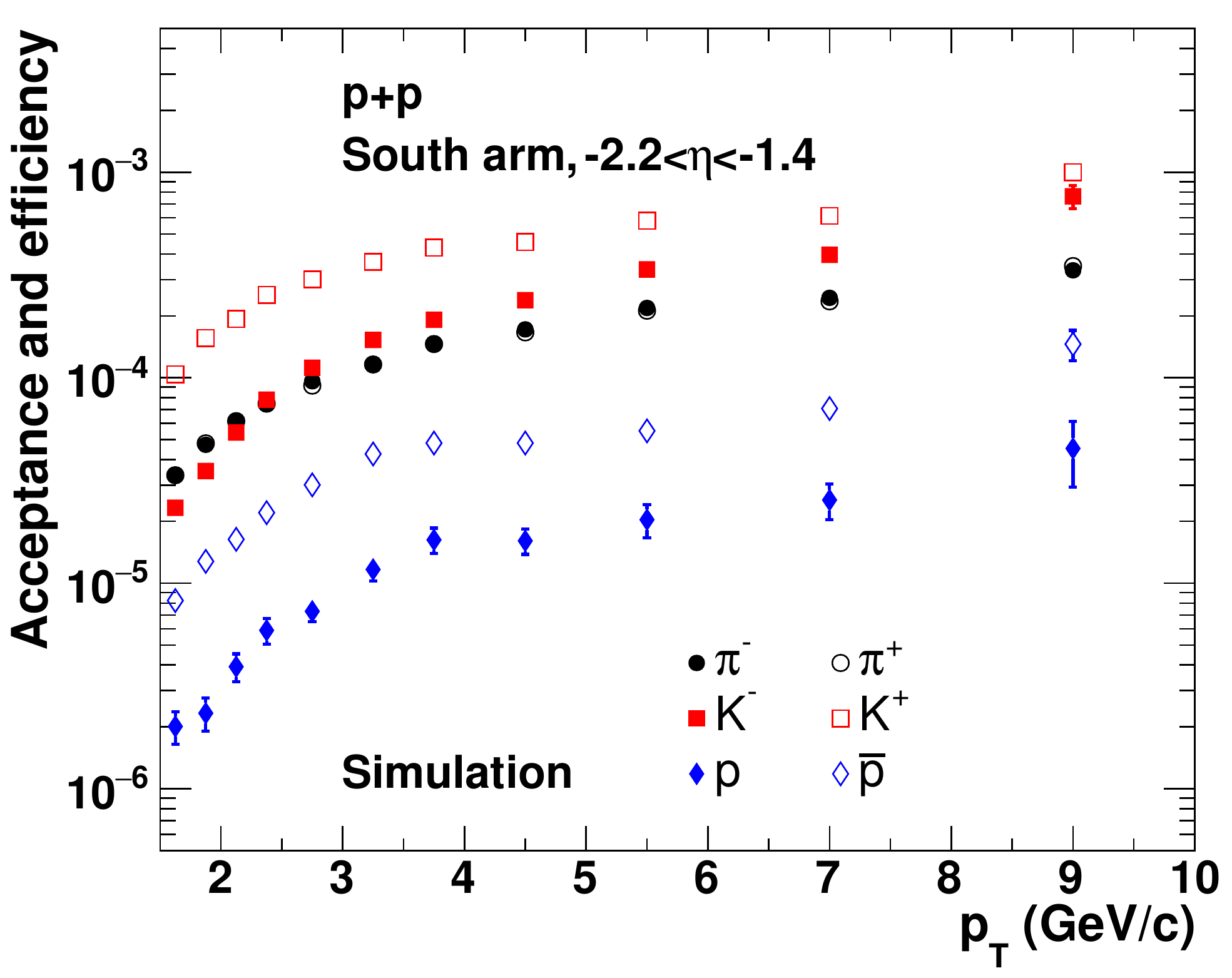}
\caption{\label{fig:acceff_pid_pp}
Acceptance and reconstruction efficiency for different charged hadrons 
as a function of \pt in the south arm evaluated in \pp collisions.}
\end{figure}

\begin{figure*}[htb]
\includegraphics[width=0.998\linewidth]{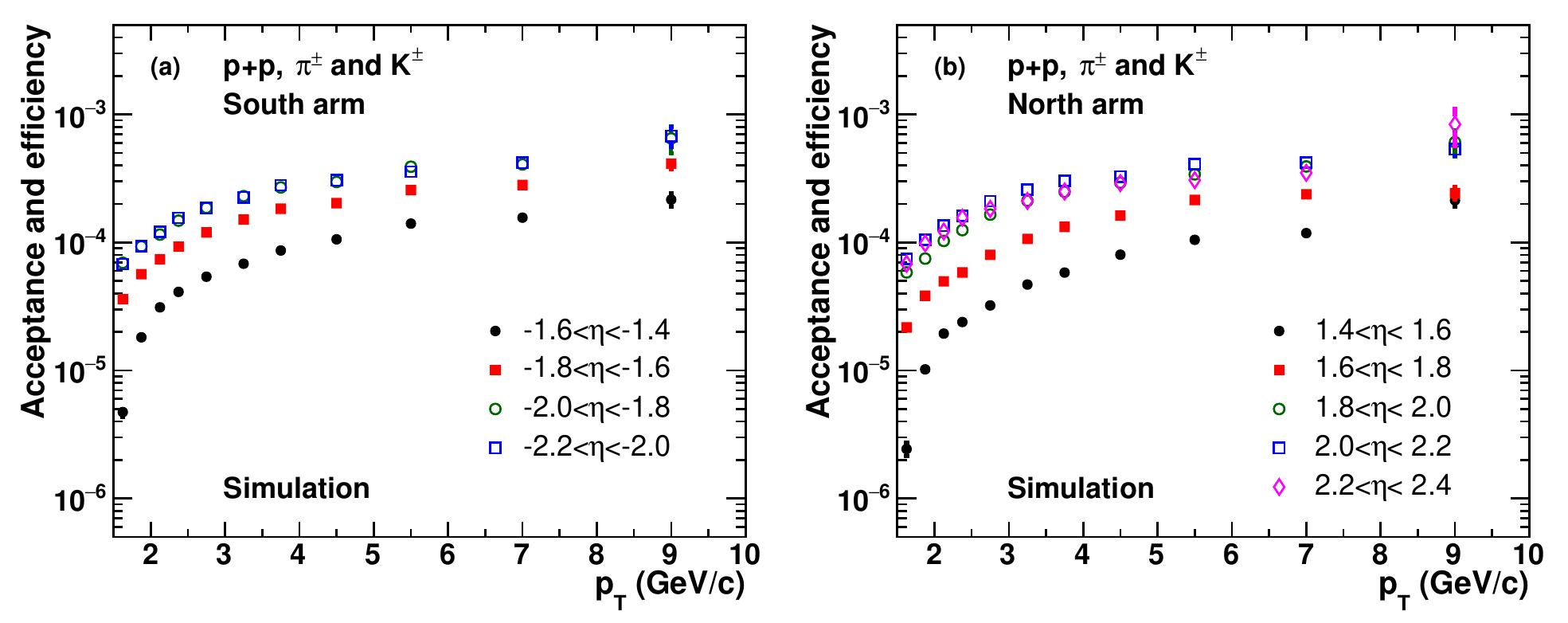}
\caption{\label{fig:acceff_etabin_pp}
Combined acceptance and reconstruction efficiency for $\pi^{\pm}$ and 
$K^{\pm}$ as a function of \pt in the (a) south and (b) north arms 
evaluated in \pp collisions.}
\end{figure*}

One consideration with the FVTX association requirement is the 
possibility of multiple FVTX tracks within the search window of a 
projected MuTr track, due to the higher FVTX track multiplicity and the 
smeared momentum information from the MuTr as shown in 
Fig.~\ref{fig:g4sim_deta} (a). In this case, a MuTr track can be 
associated with a wrong FVTX track. This is referred to as a 
mis-association. Such mis-associations result in further smearing of the 
reconstructed \pt and $\eta$.  The FVTX-MuTr association efficiency 
depends on the event multiplicity. The probability of mis-association 
can be evaluated with a data driven method developed 
in~\cite{Aidala:2017yte,Aidala:2017iad} by associating a MuTr track with 
FVTX tracks from another event of similar FVTX track multiplicity. The 
same method has been used in this analysis, and the estimated fraction 
of mis-associations in the \pp data is $\sim1.5\%$ at 
$\pt\approx1.5~{\rm GeV}/c$ and decreases down to $\sim0.5\%$ at 
$\pt\approx5~{\rm GeV}/c$. In the 0\%--5\% highest multiplicity \pau 
collisions, the estimated fraction of mis-associations in the south arm 
(Au-going direction) is $\sim3\%$ at $\pt\approx1.5~{\rm GeV}/c$ and 
$\sim1\%$ at $\pt\approx5~{\rm GeV}/c$, which is a factor of two higher 
than the estimate for \pp collisions. The mis-association fraction is 
also checked with hadron simulation events embedded into real data 
events, and is consistent with the data driven values. The embedding 
simulation described in Sec.~\ref{sec:acceff} is used to take into 
account the multiplicity dependent FVTX-MuTr association efficiency.

In addition to the requirements on \pz and FVTX-MuTr association, track 
quality cuts are applied. MuTr tracks are required to have at least 11 
hits out of a maximum of 16 hits, and a 3$\sigma$ MuTr track fit quality 
cut is applied. For association between MuTr and MuID tracks, three 
standard deviation cuts are applied to the angle and distance between 
MuTr and MuID tracks projected to the MuID gap 0. The associated FVTX 
track is required to have hits in at least three of the four stations, 
and an additional 3$\sigma$ fit quality cut is applied. 
Momentum-dependent cuts are applied to the angle difference in the 
radial and azimuthal directions between FVTX and MuTr tracks projected 
to the middle of the absorber ($z=70$~cm).  These selections help reject 
tracks from decay muons, secondary hadrons, and FVTX-MuTr 
mis-associations.

The trigger efficiency is evaluated using hadron candidates from MB 
triggered events by measuring the fraction of hadron candidates 
satisfying the trigger requirements. Figure~\ref{fig:trigeff_pp} shows 
the trigger efficiency for hadrons as a function of \pt at MuID gap 2 
and gap 3 of the south arm in the \pp data.  The trigger efficiency for 
hadrons at MuID gap 3 is higher than that for hadrons at MuID gap 2. The 
efficiency at the north arm in the \pp data is similar. Due to the 
larger statistical fluctuations at $\pt>5~{\rm GeV}/c$, a fit function 
is used to obtain the \pt-dependent trigger efficiency correction 
factors. The trigger efficiency is separately evaluated for each muon 
arm as well as each centrality bin of \palau collisions to account for 
possible multiplicity effects and detector performance variation during 
the data taking period.  The relative variation of the trigger 
efficiency over the data taking period is less than 10\%. Because this 
variation of the trigger efficiency is accounted for by the detector 
performance variation described in Sec.~\ref{sec:acceff}, no additional 
systematic uncertainty is assigned.

\subsection{Acceptance and reconstruction efficiency}
\label{sec:acceff}

Calculation of the absolute acceptance and efficiency for hadrons 
requires a detailed simulation of the hadronic interactions in the thick 
absorber material. There are significant uncertainties as observed from 
various \geant implementations of such interactions. However, the 
response of hadrons inside the absorber is independent of collision 
systems, and hence this uncertainty will cancel out when comparing 
hadron yields between two collision systems. Therefore, nuclear effects 
on hadron production can be studied by taking into account only the 
additional multiplicity-dependent efficiency corrections. To 
obtain the multiplicity-dependent efficiency corrections, a full \geant 
detector simulation was developed as follows:
\begin{enumerate}

\item Generate a mixture of hadrons ($\pi^{\pm}$, $K^{\pm}$, 
$K^{0}_{S}$, $K^{0}_{L}$, $p$, and $\bar{p}$) based on initial \pt and 
$\eta$ distributions studied in~\cite{Adare:2012px,Adare:2013lkk}. Based 
on measurements of identified charged hadrons at 
midrapidity~\cite{Adare:2011vy,Agakishiev:2011dc,Adare:2013esx}, an 
extrapolation to forward and backward rapidity is done by multiplying 
the ratio of \pt spectra between mid and forward/backward rapidity from 
event generators~\cite{Sjostrand:2006za,Gyulassy:1994ew}. These 
simulated hadrons originate from a $z$ distribution which matches the 
measured \zbbc data.

\item Run a full \geant simulation for the detector response of hadrons.

\item Reconstruct simulated detector hits embedded on top of background 
hits from real data for each centrality bin in each collision system.  
Apply the data-driven detector dead channel maps to account for 
variations in detector performance.

\end{enumerate}

Figure~\ref{fig:acceff_pid_pp} shows an example of acceptance and 
efficiency result as a function of \pt for different species of hadrons 
at MuID gaps 2 and 3 of the south arm in \pp collisions. The acceptance 
and efficiency for $\pi^{\pm}$ and $K^{-}$ is comparable, and $K^{+}$ 
has the highest acceptance and efficiency due to its longer nuclear 
interaction length. The acceptance and efficiency for $p$ and $\bar{p}$ 
is much smaller than other charged hadrons.

Due to these species-dependent corrections, the overall acceptance and 
efficiency will depend on the relative production of these hadrons. In 
order to correctly account for the species dependence, an initial 
$K^{\pm}/\pi^{\pm}$ ratio for each collision system is estimated 
separately. The contribution of $p$ and $\bar{p}$ to reconstructed 
tracks based on this hadron simulation is less than 5\%, and thus we do 
not include them in the overall result. 
Figure~\ref{fig:acceff_etabin_pp} shows the combined acceptance and 
efficiency for $\pi^{\pm}$ and $K^{\pm}$ as a function of \pt in \pp 
collisions for various $\eta$ ranges. The acceptance and efficiency is 
higher at more forward rapidity where path length through the absorber 
is shorter, and the total momentum of tracks for a given \pt range is 
also larger.  To have a more accurate correction, the full \pt 
and $\eta$ dependent correction is applied.

\subsection{Nuclear modification factor}

Nuclear effects on charged hadron production in \pal and \pau collisions 
are quantified with the nuclear modification factor,
\begin{equation}
R_{pA} = \frac{dY^{pA}/dp_{T}d\eta}{dY^{pp}/dp_{T}d\eta} \cdot \frac{1}{\ncoll},
\end{equation}
where $dY^{pA}/dp_{T}d\eta$ is the charged hadron yield in a certain 
centrality bin of \palau collisions.  These yields are corrected for the 
trigger efficiency, acceptance and reconstruction efficiency, and 
centrality bias factor introduced in Sec.~\ref{sec:experiment}. 
$dY^{pp}/dp_{T}d\eta$ is the hadron yield in \pp collisions corrected 
for the trigger efficiency, acceptance and reconstruction efficiency, 
and BBC efficiency. Finally \ncoll is the mean number of binary 
collisions for the corresponding centrality bin as calculated with the 
MC Glauber framework~\cite{Loizides:2014vua}. The \ncoll values, bias 
correction factors, and related systematic uncertainties for each 
centrality bin of \pal and \pau collisions appear in 
Table~\ref{tab:centrality}.

\begin{table}[tbh]
\caption{\label{tab:centrality}
The \ncoll and centrality bias correction factors are shown for 
different centrality selections of \pal and \pau collisions.
}
\begin{ruledtabular} \begin{tabular}{cccc}
 collision system & centrality & \ncoll & bias factor \\\hline
 \pal & 0\%--5\%   & 4.1$\pm$0.4  & 0.75$\pm$0.01 \\
      & 5\%--10\%  & 3.5$\pm$0.3  & 0.81$\pm$0.01 \\
      & 10\%--20\% & 2.9$\pm$0.3  & 0.84$\pm$0.01 \\
      & 20\%--40\% & 2.4$\pm$0.1  & 0.90$\pm$0.02 \\
      & 40\%--72\% & 1.7$\pm$0.1  & 1.04$\pm$0.04 \\  
      & 0\%--100\% & 2.1$\pm$0.1  & 0.80$\pm$0.02 \\
\\
 \pau & 0\%--5\%   & 9.7$\pm$0.6  & 0.86$\pm$0.01 \\
      & 5\%--10\%  & 8.4$\pm$0.6  & 0.90$\pm$0.01 \\
      & 10\%--20\% & 7.4$\pm$0.5  & 0.94$\pm$0.01 \\
      & 20\%--40\% & 6.1$\pm$0.4  & 0.98$\pm$0.01 \\
      & 40\%--60\% & 4.4$\pm$0.3  & 1.03$\pm$0.01 \\
      & 60\%--84\% & 2.6$\pm$0.2  & 1.00$\pm$0.06 \\
      & 0\%--100\% & 4.7$\pm$0.3  & 0.86$\pm$0.01 \\
\end{tabular} \end{ruledtabular}
\end{table}

\section{Systematic uncertainties}
\label{sec:sys_uncert}

In this section, sources of systematic uncertainty in the nuclear 
modification factor are described, and the procedure used to determine 
each systematic uncertainty is discussed.

\subsection{Acceptance and efficiency}

\subsubsection{Initial hadron distribution}

Because there are limited measurements of identified charged hadrons at 
forward and backward rapidity ($1.2<|\eta|<2.4$), some model assumptions 
are necessary.  Such forward rapidity particle yields have previously 
been estimated for use in earlier PHENIX \pp and \dau collisions 
studies---see Ref.~\cite{Adare:2013lkk} for details.  Here we follow 
that previous work as input for our simulation studies. To 
account for uncertainties on the estimated \pt and $\eta$ distributions, 
weight factors in \pt and $\eta$ for each collision system are 
extracted by comparing reconstructed \pt and $\eta$ distributions 
between data and simulation. The variation of acceptance and efficiency 
with modified initial \pt and $\eta$ distributions based on the 
weighting factors is less than 3\% for \pp data. For \palau data, the 
variation at forward (backward) rapidity is less than 3\% (5\%). The 
variation is included in the systematic uncertainty.

In addition, there is an uncertainty in the $K^{\pm}/\pi^{\pm}$ ratio 
which influences the combined acceptance and efficiency due to the 
longer nuclear interaction length of $K^{+}$. Based on the uncertainties 
of measurements at 
midrapidity~\cite{Adare:2011vy,Agakishiev:2011dc,Adare:2013esx} used as 
an input for extrapolation to forward and backward rapidity and a 
possible extrapolation uncertainty estimated by comparing with the data 
at more forward rapidity~\cite{Arsene:2007jd}, an effect of a $\pm$30\% 
variation of $K^{\pm}/\pi^{\pm}$ on the acceptance and efficiency has 
been evaluated.

The $K^{\pm}/\pi^{\pm}$ at midrapidity in various centrality bins of 
\dau collisions are compatible with each other~\cite{Adare:2013esx}, and 
the difference of $K^{\pm}/\pi^{\pm}$ between \dau and \palau collisions 
in \hijing~\cite{Gyulassy:1994ew} is less than 10\%. These additional 
sources of uncertainty are covered by the 30\% variation of 
$K^{\pm}/\pi^{\pm}$. The variation of acceptance and efficiency due to 
the 30\% $K^{\pm}/\pi^{\pm}$ change is less than 5\% (7\%) in \pp 
(\palau) collisions.

\subsubsection{Proton contamination}

As described in Sec.~\ref{sec:acceff}, the acceptance and efficiency is 
calculated for $\pi^{\pm}$ and $K^{\pm}$. There is an $\sim5\%$ proton 
contamination where the fraction may vary with the initial $p/(\pi+K)$ 
ratio. Based on the results in \pp and \dau collisions at 
midrapidity~\cite{Adare:2011vy,Agakishiev:2011dc,Adare:2013esx}, the 
$p/\pi$ ratio at $\pt\approx2~{\rm GeV}/c$ in 0\%--20\% central \dau 
collisions is about 30\% larger than in \pp collisions, which results in 
an increase of the contamination to 6.5\% in 0\%--20\% central \dau 
collisions as compared with 5\% in \pp collisions. However, there is a 
lack of $p/\pi$ measurements in a broader \pt range in various 
centrality ranges of \palau collisions. Therefore, a conservative 
uncertainty of 5\% is assigned corresponding to a factor of two 
difference in $p/(\pi+K)$ ratios between \palaup collisions.

\subsubsection{Hadron simulation}

Although hadron response inside the absorber will not vary between 
different collision systems, the variation of acceptance and efficiency 
among three hadron interaction models (\textsc{qgsp bert}, \textsc{qgsp 
bic}, and \textsc{ftfp bert}) in \geant has been checked. A detailed 
description of the three models and a previous study for muons can be 
found in Refs.~\cite{Agostinelli:2002hh,Aidala:2018ajl}. The variation 
of the combined acceptance and efficiency for $\pi^{\pm}$ and $K^{\pm}$ 
between the three models is less than 2\% in \pt and $\eta$.

\subsubsection{Variation of detector efficiency}

During the data taking period, the detector performance varied due to 
temporary dead channels, changes in the instantaneous beam luminosity, 
and other experimental factors. The average detector efficiency for each 
collision system is included in the hadron simulation.  The raw yield 
variation in FVTX and muon tracks is considered as a source of 
systematic uncertainty. The level of variation appears in 
Table~\ref{tab:sys_detector}. The FVTX performance is quite stable 
during the entire data taking period, and the variation of the muon arm 
is observed to be larger in the south arm in the \pau data due to a 
larger sensitivity of the MuID efficiency to the instantaneous beam 
luminosity of Au ions.  A 1$\sigma$ variation of the raw yield is 
assigned as a systematic uncertainty for each detector, and two 
systematic uncertainties are added in quadrature.

\begin{table}[tbh]
\caption{\label{tab:sys_detector}
Variation of detector performance for the south (S) and north (N)
muon-arm spectrometers, as characterized by the number of FVTX 
and MuTr-MuID tracks per event.}
\begin{ruledtabular}
\begin{tabular}{ccc}
 Collision system & FVTX & MuTr-MuID \\\hline
 \pp & 2.8\%(S), 2.6\%(N) & 4.8\%(S), 5.6\%(N) \\
 \pal & 2.4\%(S), 2.1\%(N) & 3.0\%(S), 2.8\%(N) \\
 \pau & 2.7\%(S), 2.3\%(N) & 7.2\%(S), 2.7\%(N) \\
\end{tabular}
\end{ruledtabular}
\end{table}

\subsubsection{FVTX-MuTr mis-association}

The probability of FVTX-MuTr mis-association depends on the FVTX track 
multiplicity, and the mis-association may artificially increase the 
acceptance and efficiency when requiring FVTX track association. The 
procedure for calculating the acceptance and efficiency using embedded 
simulations takes into account the multiplicity dependent FVTX-MuTr 
mis-association. The primary method to estimate the fraction of 
FVTX-MuTr is the data driven method described in 
Sec.~\ref{sec:hadronsel}, and the systematic uncertainty is evaluated by 
comparing with the estimated fraction from the embedded simulation. The 
difference is less than 1\% of the maximum $\sim3\%$ of FVTX-MuTr 
mis-association contamination in 0\%--5\% \pau collisions. A 1\% 
systematic uncertainty is assigned for the estimation of FVTX-MuTr 
mis-association.

\subsubsection{Vertex resolution}

Because the location of the FVTX is close to the interaction point, the 
$\eta$ acceptance of the FVTX depends on the $z$ position of collisions. 
In the hadron simulation for acceptance and efficiency calculation, the 
measured \zbbc distribution for each collision system is used, but there 
is uncertainty due to the resolution of \zbbc. When considering the 2~cm 
of \zbbc resolution, the variation of acceptance and efficiency is less 
than 0.5\% in all three collision systems. A 0.5\% systematic 
uncertainty is assigned due to the \zbbc resolution.

\subsection{Contamination from secondary hadrons}

Remaining secondary hadrons can introduce a smearing of kinematic 
variables (\pt and $\eta$) used in this analysis. The hadron simulation 
for calculating acceptance and efficiency already includes this 
component, however there can be a discrepancy in the relative 
contribution of secondary hadrons between the data and simulation. The 
systematic uncertainty on \rpa is estimated by varying the FVTX-MuTr 
matching quality cuts (projection angles between FVTX and MuTr tracks) 
which affect the remaining fraction of secondary hadrons. Based on the 
hadron simulation, a tighter or looser FVTX-MuTr matching quality cut 
changes the relative fraction of secondary hadrons by $\sim25\%$; the 
variation on \rpa is less than 3\%.

\subsection{Multiple collisions}

Due to the high instantaneous beam luminosity particularly in \pp and 
\pal collisions, there is a chance of having multiple collisions in a 
single bunch crossing. This can introduce a bias in the yield 
calculation as well as centrality determination. The effect has been 
checked by analyzing two data groups with low and high instantaneous 
beam luminosity, and the difference in \rpa is less than 5\%. The 
variation due to multiple collisions is already considered in the 
systematic uncertainty from the variations in detector efficiency with 
data-taking period. Therefore, no additional systematic uncertainty is 
assigned.

\subsection{BBC efficiency and centrality selection}

The BBC efficiency in \pp collisions is $\sim55\%$ for MB events and 
$\sim79\%$ for hard scattering events, and a 10\% systematic uncertainty 
is assigned based on previous studies~\cite{Adler:2003pb}.  This 
uncertainty is a global scale uncertainty.

As described in Table~\ref{tab:centrality}, there are systematic 
uncertainties on \ncoll and bias correction factor calculations. The 
procedure to estimate these systematic uncertainties has been studied 
for \dau collisions~\cite{Adare:2013nff}, and the same procedure is used 
for \pal and \pau collisions.

\subsection{Summary of systematic uncertainty}

Table~\ref{tab:sys_summary} shows the summary of systematic 
uncertainties. All systematic uncertainties are point-to-point 
correlated. Because most of sources on the acceptance and efficiency are 
independent in each collision system, there is no cancellation of 
systematic uncertainty for \rpa calculation.

\begin{table}[tbh]
\caption{\label{tab:sys_summary}
Summary of systematic uncertainties.}
\begin{ruledtabular} \begin{tabular}{cc}
 Source & Relative uncertainty \\\hline
 & 9.5--9.9\% (\pp) \\
Acceptance and efficiency & 9.8--10.7\% (\pal) \\
 & 9.8--12.6\% (\pau) \\
\\
 Secondary hadron & 3\%\\
\\
\multirow{3}{*} {\minitab[c]{BBC efficiency and \\ 
centrality bias correction}} & 10\% (\pp)\\
  & 1.3--4.2\% (\pal)\\
  & 0.4--1.2\% (\pau)\\
\\
\multirow{2}{*}{\ncoll} & 4.7--8.5\% (\pal)\\
 & 5.8--6.6\% (\pau)\\
\end{tabular} \end{ruledtabular}
\end{table}

\begin{table*}
\caption{\label{tab:pythia}
Parameter used in \textsc{pythia8}}
\begin{ruledtabular} \begin{tabular}{ccc}
 parameter & value & description\\\hline
 SoftQCD:inelastic=on & on & QCD process for MB\\
 PDF:pSet & 7 & \cteq parton distribution function \\
 MultipartonInteractions:Kfactor & 0.5 &
 Multiplication factor for multiparton interaction\\
\end{tabular} \end{ruledtabular}
\end{table*}

\begin{figure}[htb]
\includegraphics[width=1.0\linewidth]{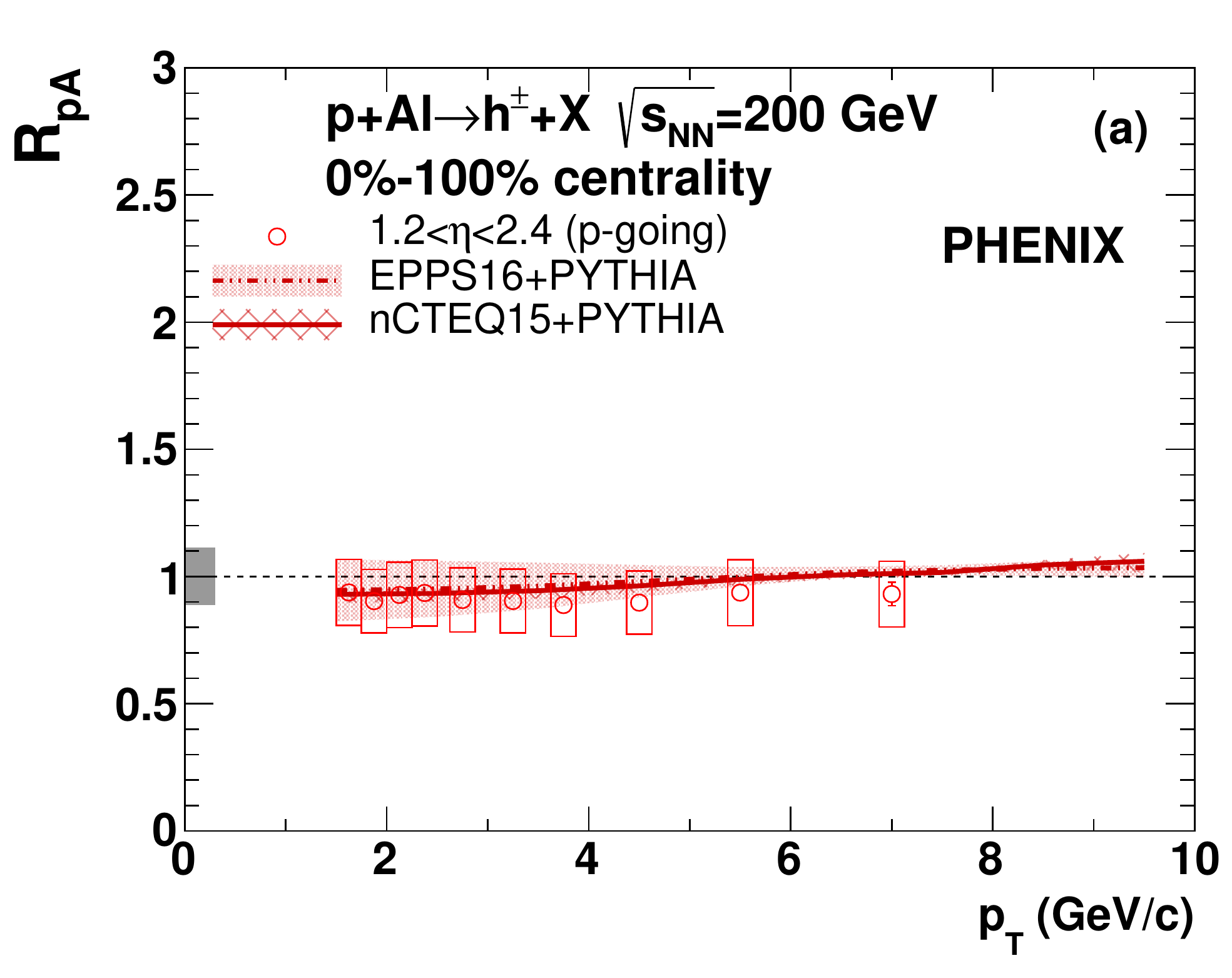}
\includegraphics[width=1.0\linewidth]{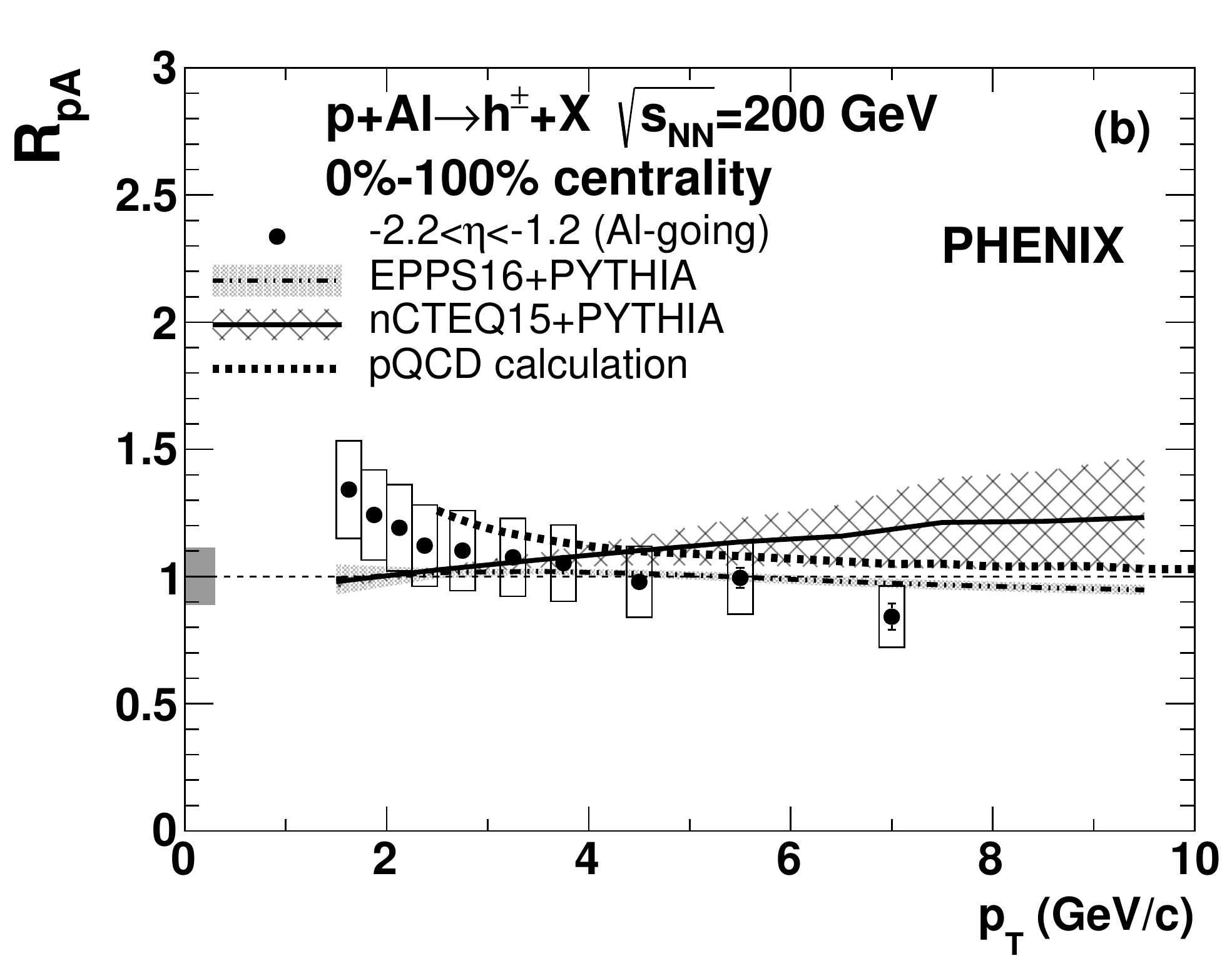}
\caption{\label{fig:rpa_pT_pal_mb} 
\rpa of charged hadrons as a function of \pt at (a) forward and 
(b) backward rapidity in \pal 0\%--100\% centrality selected 
collisions at \sqsntwo.  Also shown are comparisons to a pQCD 
calculation~\cite{Kang:2014hha} and calculations based on the nPDF 
sets~\cite{Kovarik:2015cma,Eskola:2016oht}.}
\end{figure}

\begin{figure}[htb]
\includegraphics[width=1.0\linewidth]{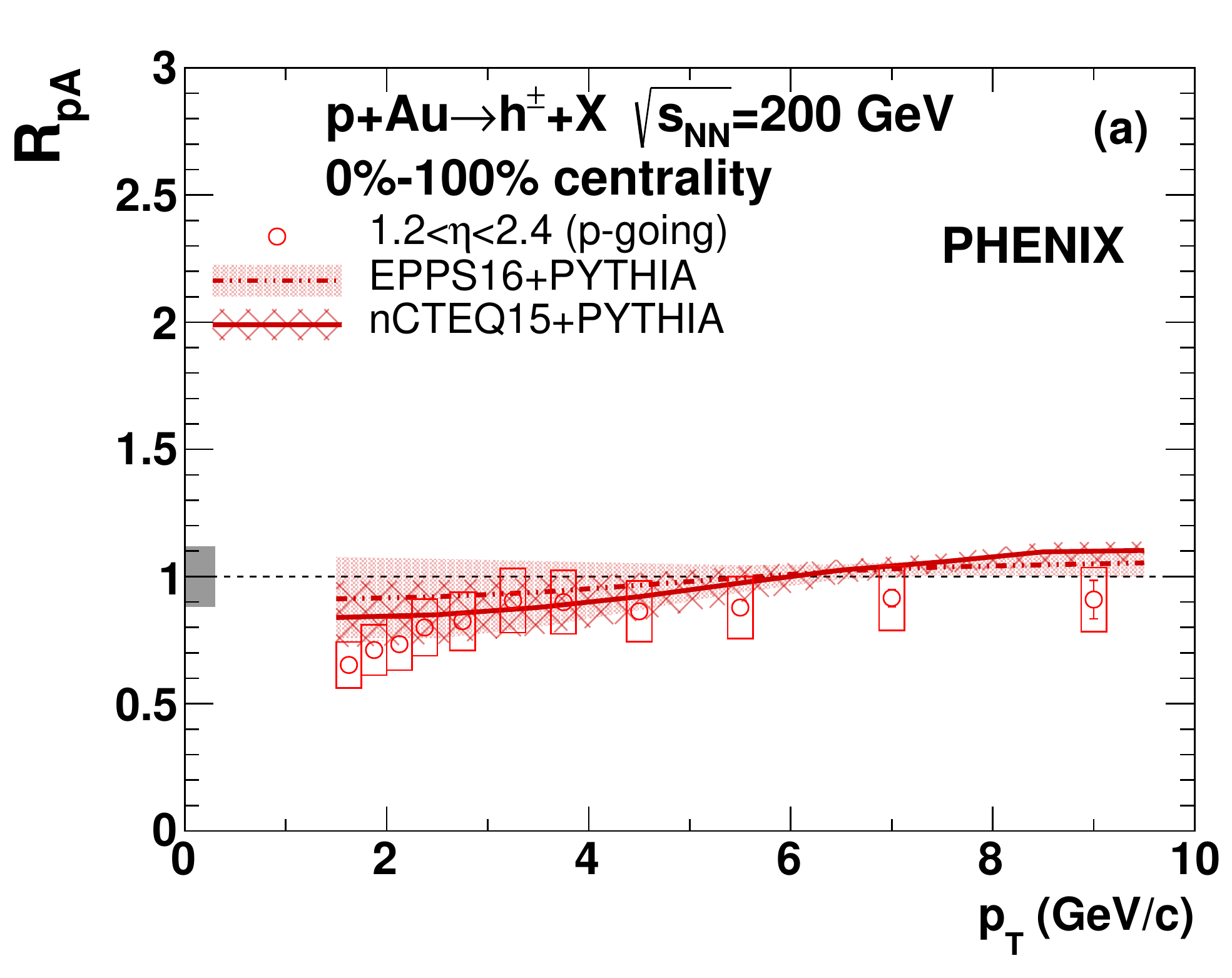}
\includegraphics[width=1.0\linewidth]{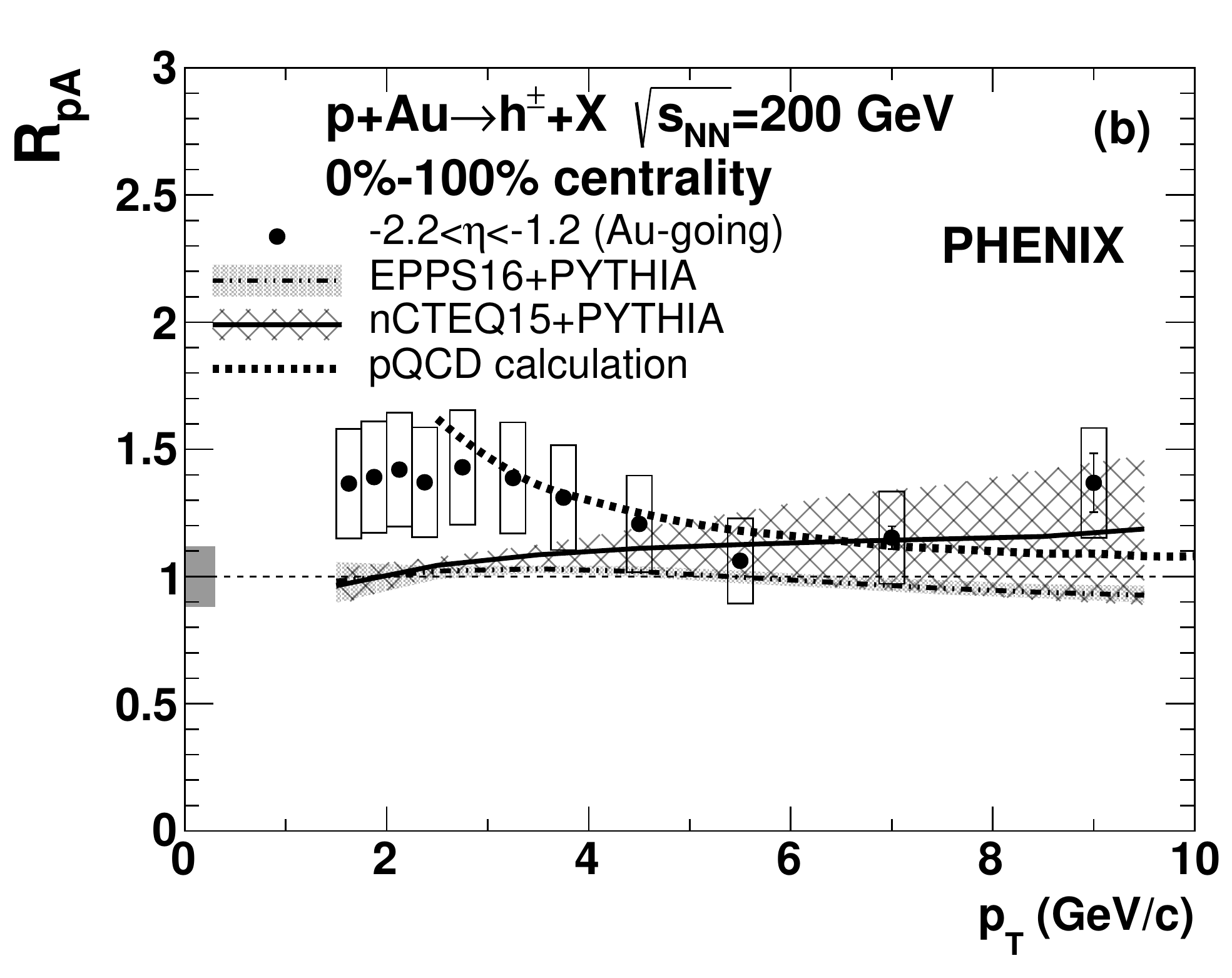}
\caption{\label{fig:rpa_pT_pau_mb} 
\rpa of charged hadrons as a function of \pt at (a) forward and 
(b) backward rapidity in \pau 0\%--100\% centrality selected 
collisions at \sqsntwo.  Also shown are comparisons to a pQCD 
calculation~\cite{Kang:2014hha} and calculations based on the nPDF 
sets~\cite{Kovarik:2015cma,Eskola:2016oht}.}
\end{figure}

\section{Results and Discussion}
\label{sec:results}

\begin{figure}[tbh]
\includegraphics[width=1.00\linewidth]{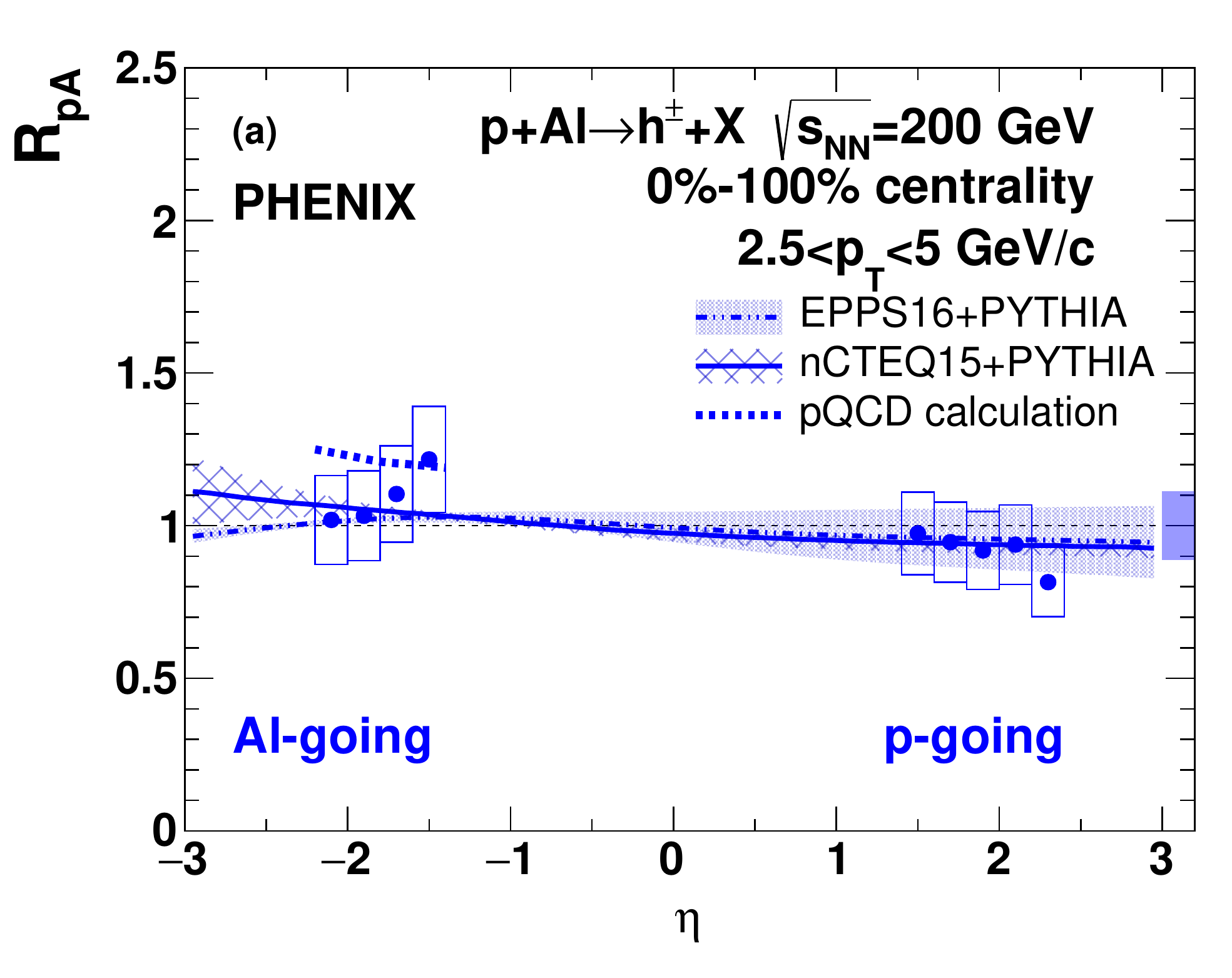}
\includegraphics[width=1.00\linewidth]{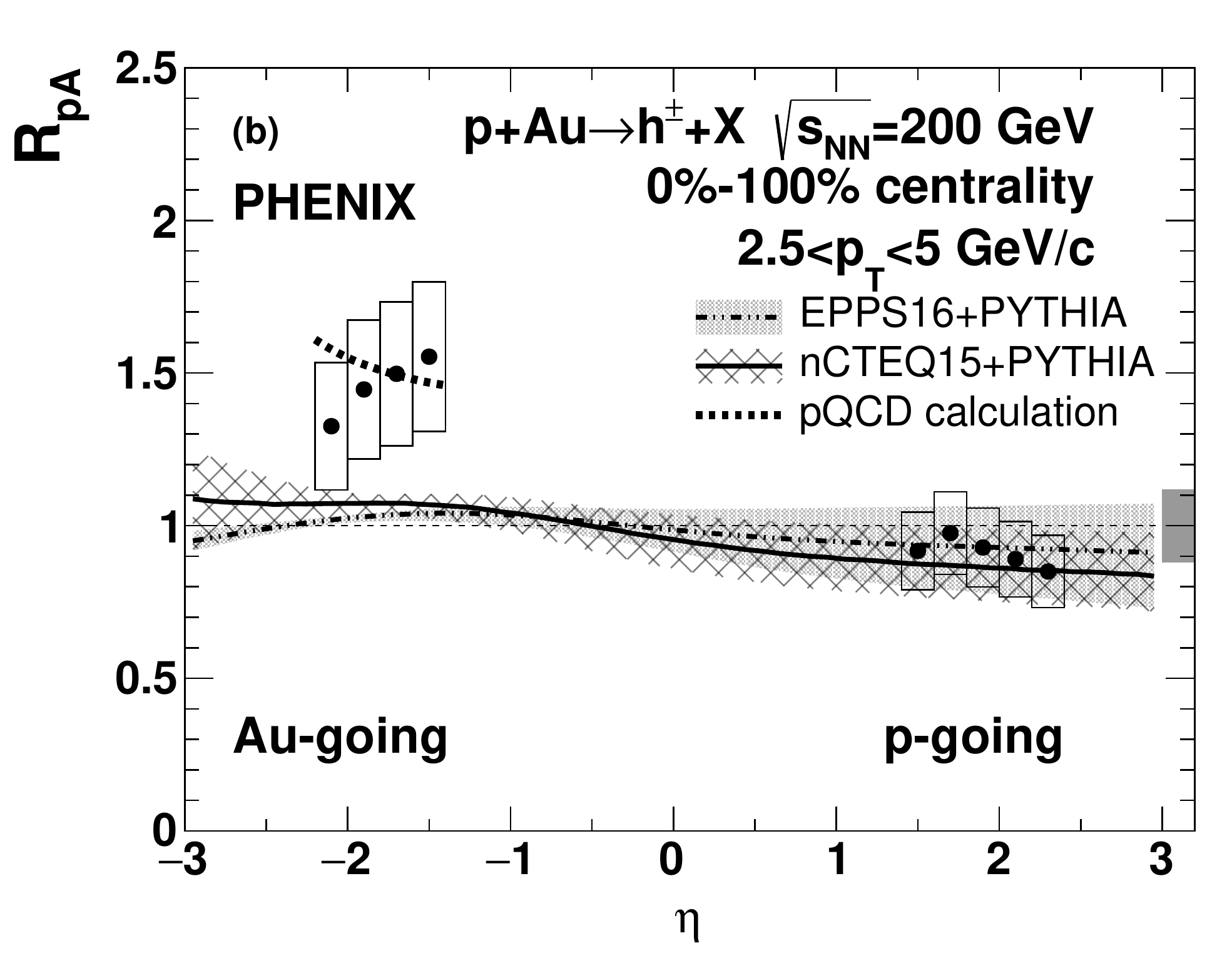}
\caption{\label{fig:rpa_eta_mb}
\rpa of charged hadrons in $2.5<p_T<5~{\rm GeV}/c$ as a function of 
$\eta$ in (a) \pal and (b) \pau 0\%--100\% centrality 
selected collisions at \sqsntwo.  Also shown are comparisons to a pQCD 
calculation~\cite{Kang:2014hha} and calculations based on the nPDF 
sets~\cite{Kovarik:2015cma,Eskola:2016oht}.}
\end{figure}

\begin{figure*}[tbh]
\includegraphics[width=0.998\linewidth]{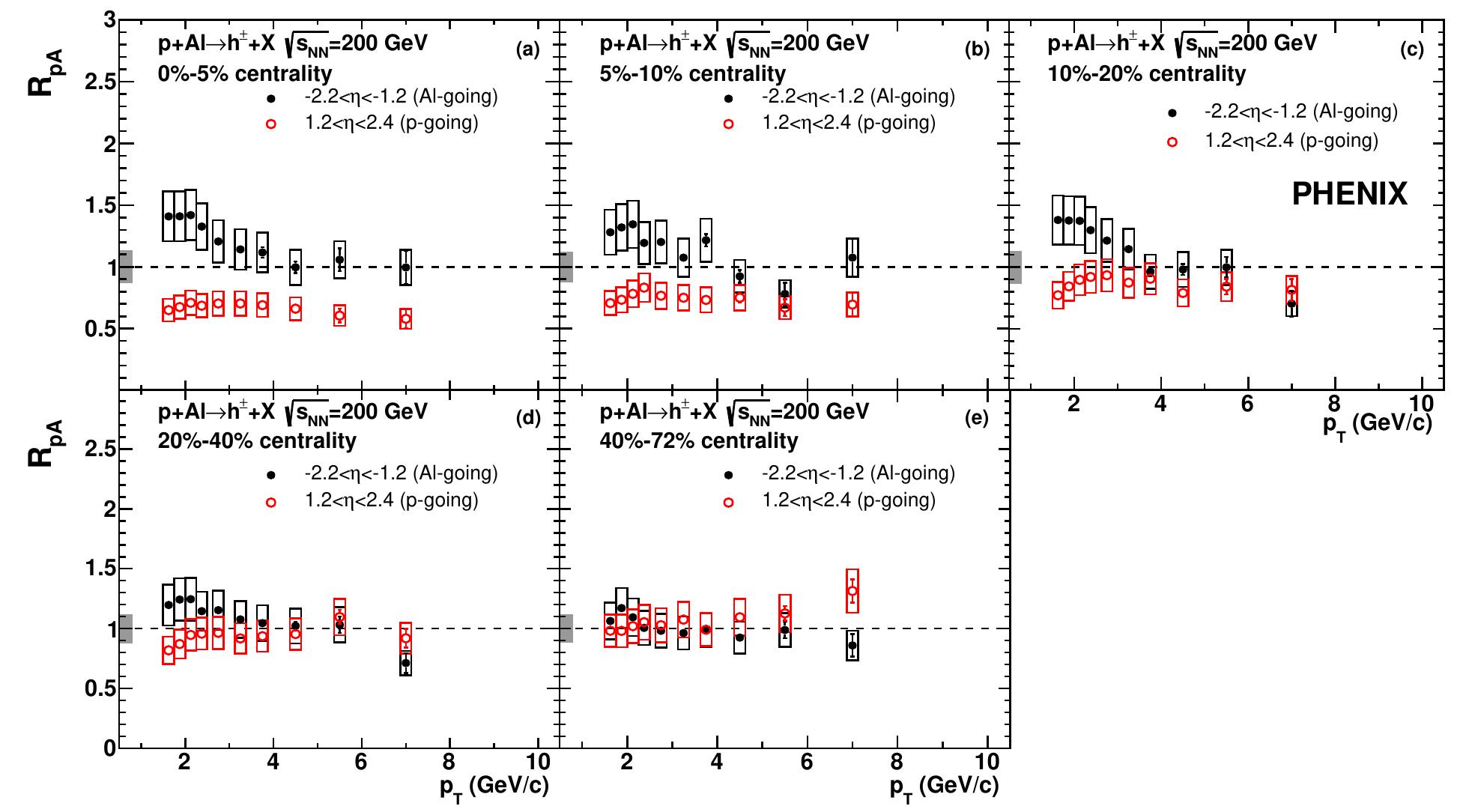}
\caption{\label{fig:rpa_pT_cent_pal}
\rpa of charged hadrons as a function of \pt at backward rapidity, 
$-2.2<\eta<-1.2$, Al-going (filled [black] circles) and forward rapidity, 
$1.4<\eta<2.4$, $p$-going (open [red] circles) in various centrality 
classes of \pal collisions at \sqsntwo.}
\vspace*{\floatsep}
\includegraphics[width=0.998\linewidth]{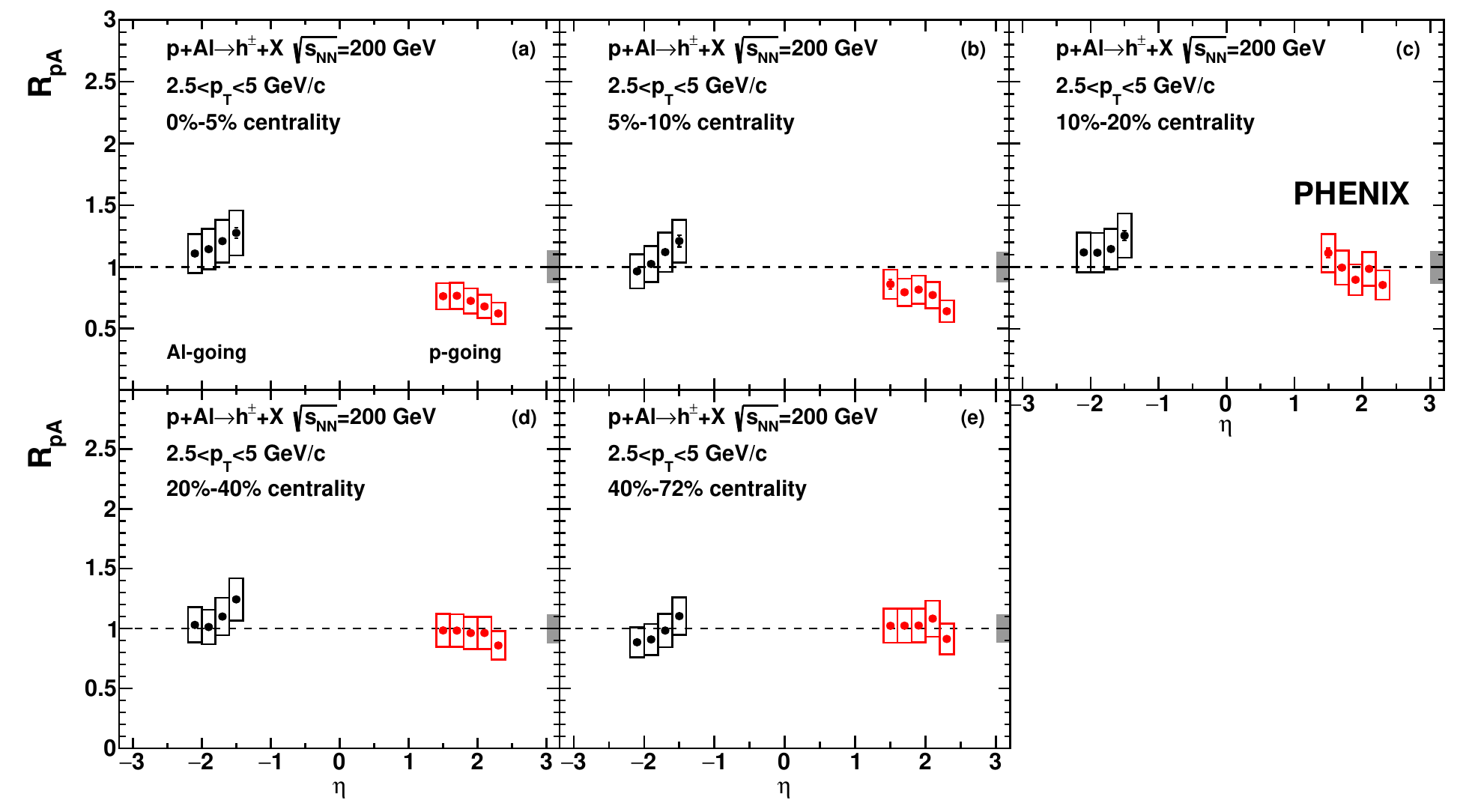}
\caption{\label{fig:rpa_eta_cent_pal}
\rpa of charged hadrons in $2.5<p_T<5~{\rm GeV}/c$ as a function of 
$\eta$ in various centrality classes of \pal collisions at \sqsntwo.}
\end{figure*}

\begin{figure*}[tbh]
\includegraphics[width=0.998\linewidth]{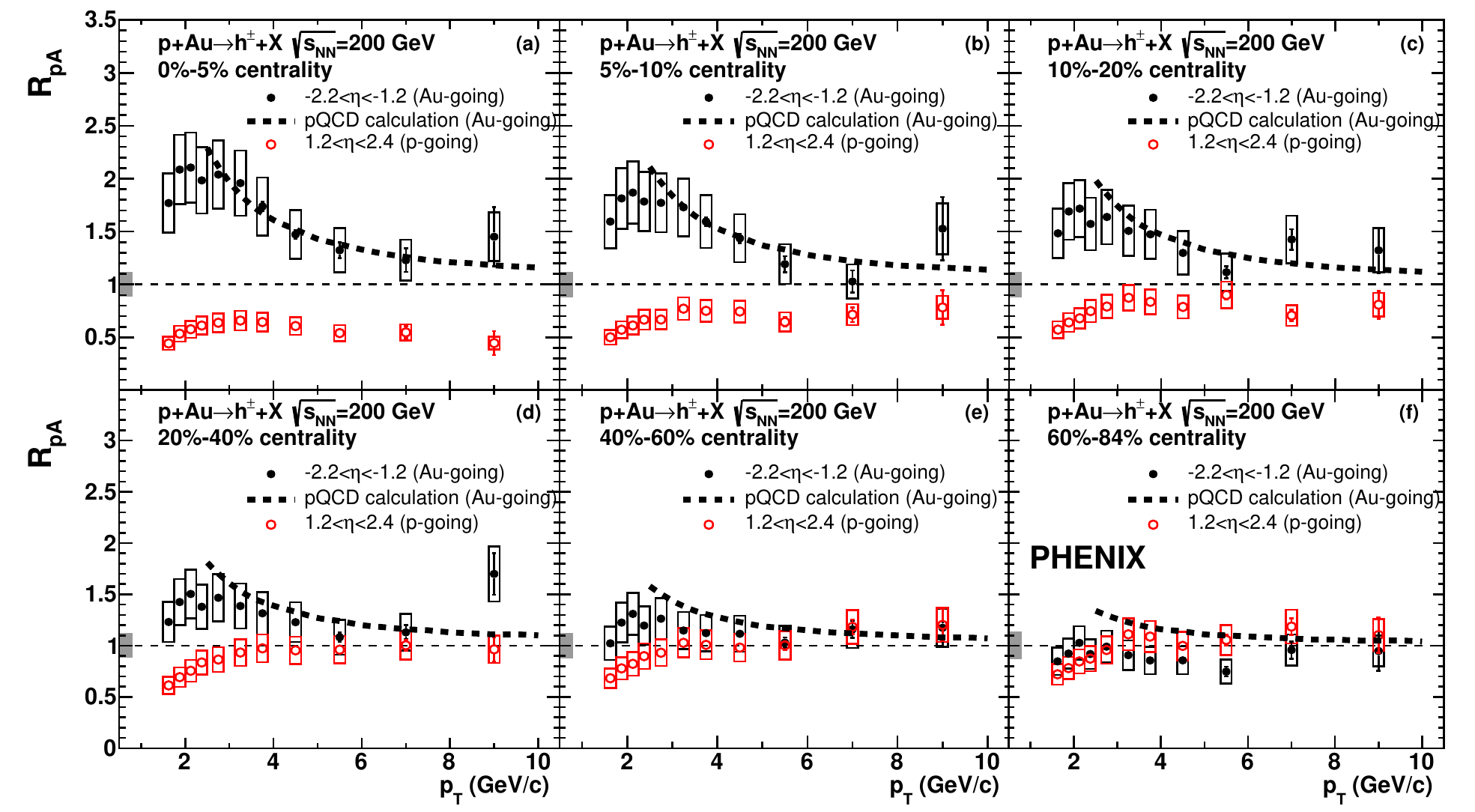}
\caption{\label{fig:rpa_pT_cent_pau}
\rpa of charged hadrons as a function of \pt at backward rapidity, 
$-2.2<\eta<-1.2$, Au-going (filled [black] circles) and forward rapidity, 
$1.4<\eta<2.4$, $p$-going (open [red] circles) in various centrality 
classes of \pau collisions at \sqsntwo.}
\vspace*{\floatsep}
\includegraphics[width=0.998\linewidth]{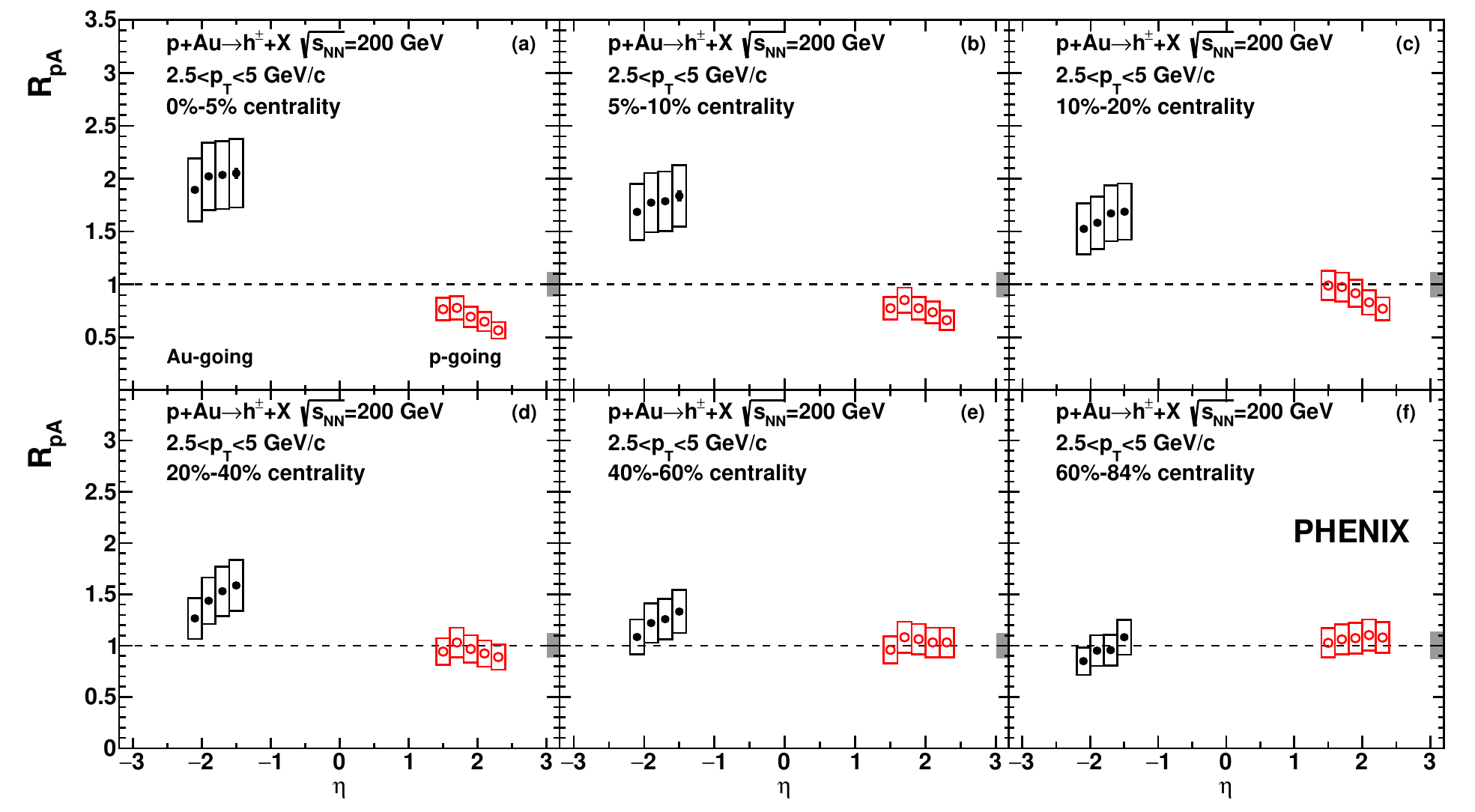}
\caption{\label{fig:rpa_eta_cent_pau}
\rpa of charged hadrons in $2.5<p_T<5~{\rm GeV}/c$ as a function of 
$\eta$ in various centrality classes of \pau collisions at \sqsntwo. 
Also shown are comparisons to a pQCD calculation~\cite{Kang:2014hha}.}
\end{figure*}

\begin{figure}[tbh]
\includegraphics[width=1.00\linewidth]{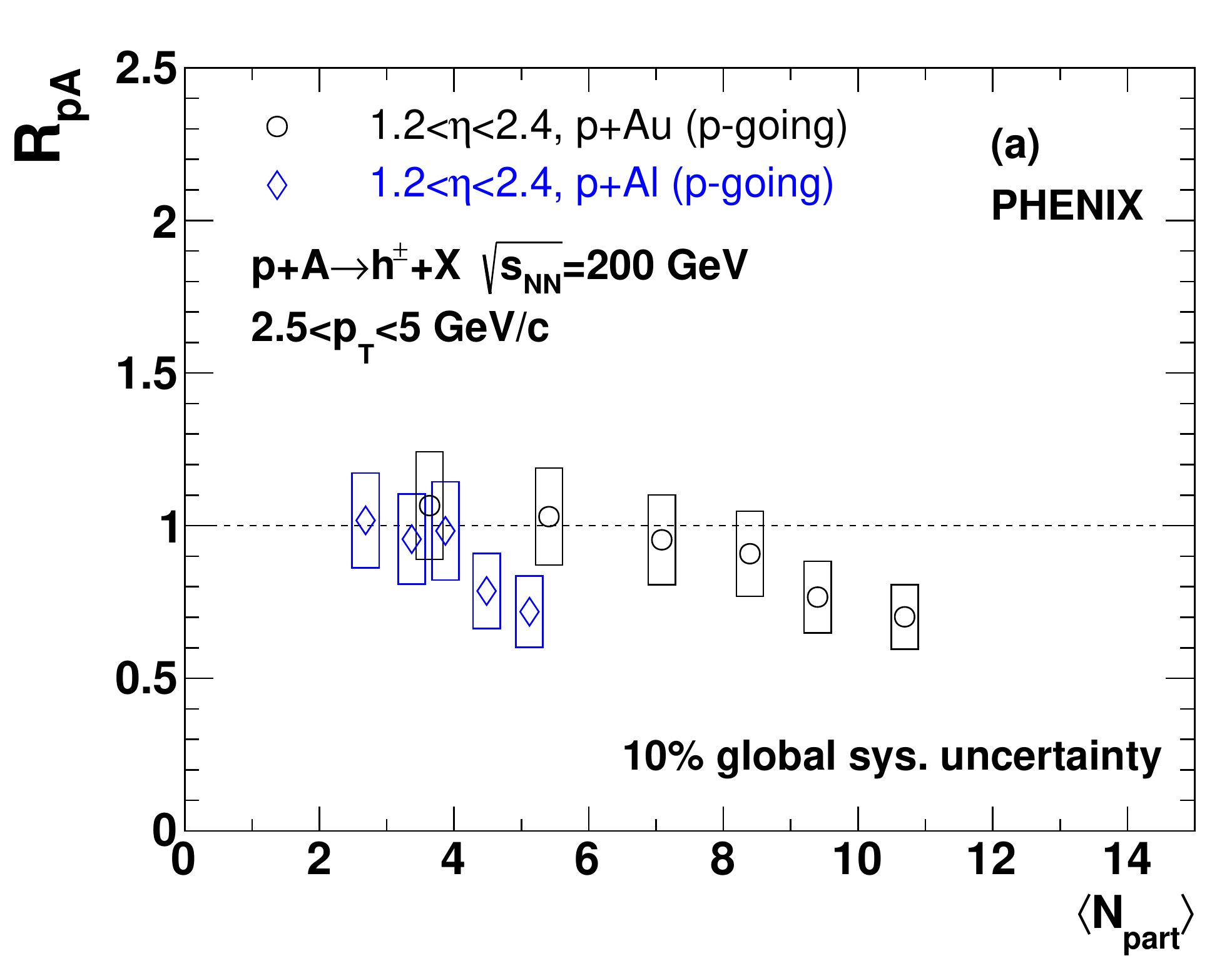}
\includegraphics[width=1.00\linewidth]{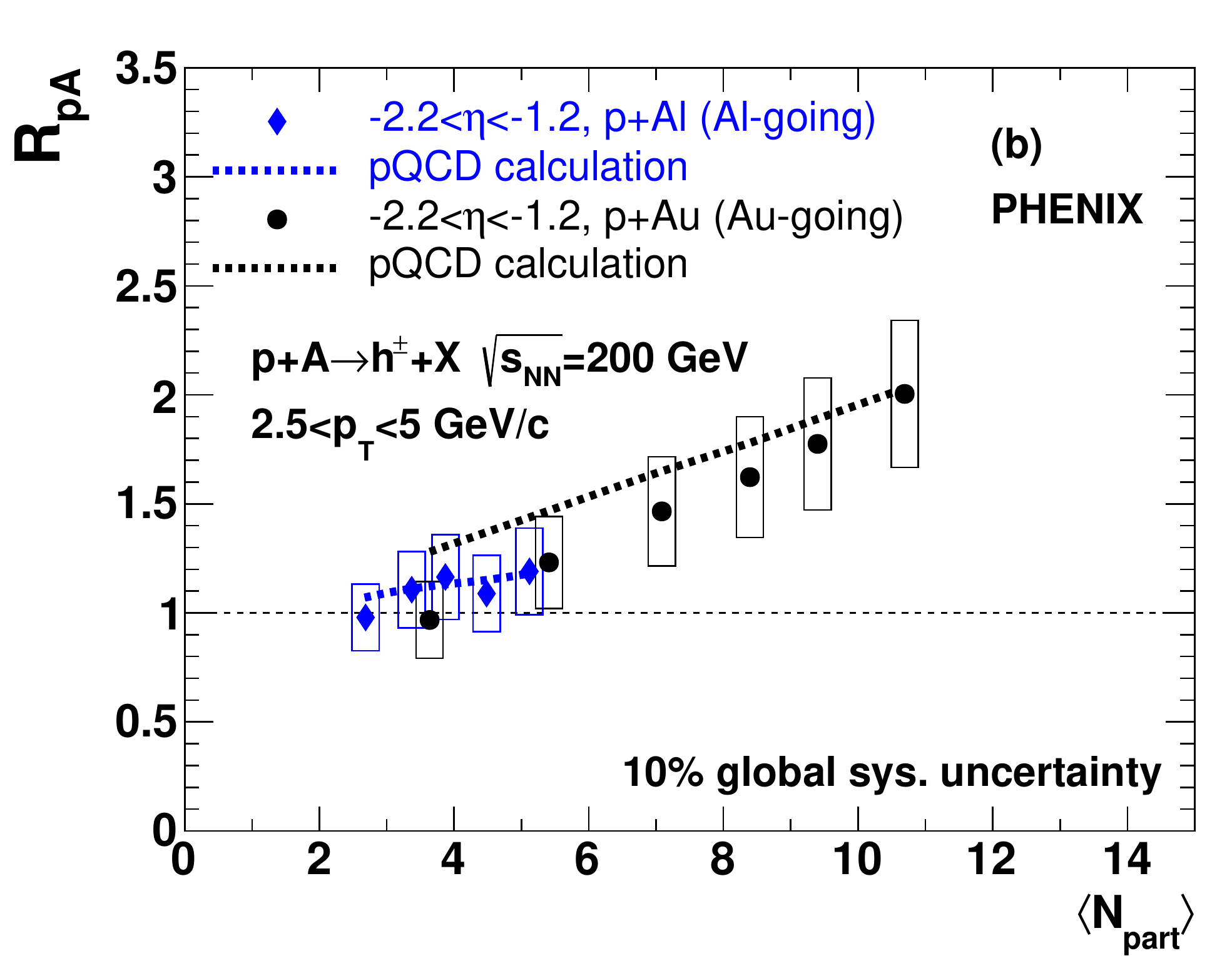}
\caption{\label{fig:rpa_npart}
\rpa of charged hadrons in $2.5<p_T<5~{\rm GeV}/c$ as a function of 
\npart at forward and backward rapidity in \pal and \pau collisions at 
\sqsntwo. Also shown are comparisons to a pQCD 
calculation~\cite{Kang:2014hha}.}
\end{figure}

\begin{figure}[tbh]
\includegraphics[width=1.00\linewidth]{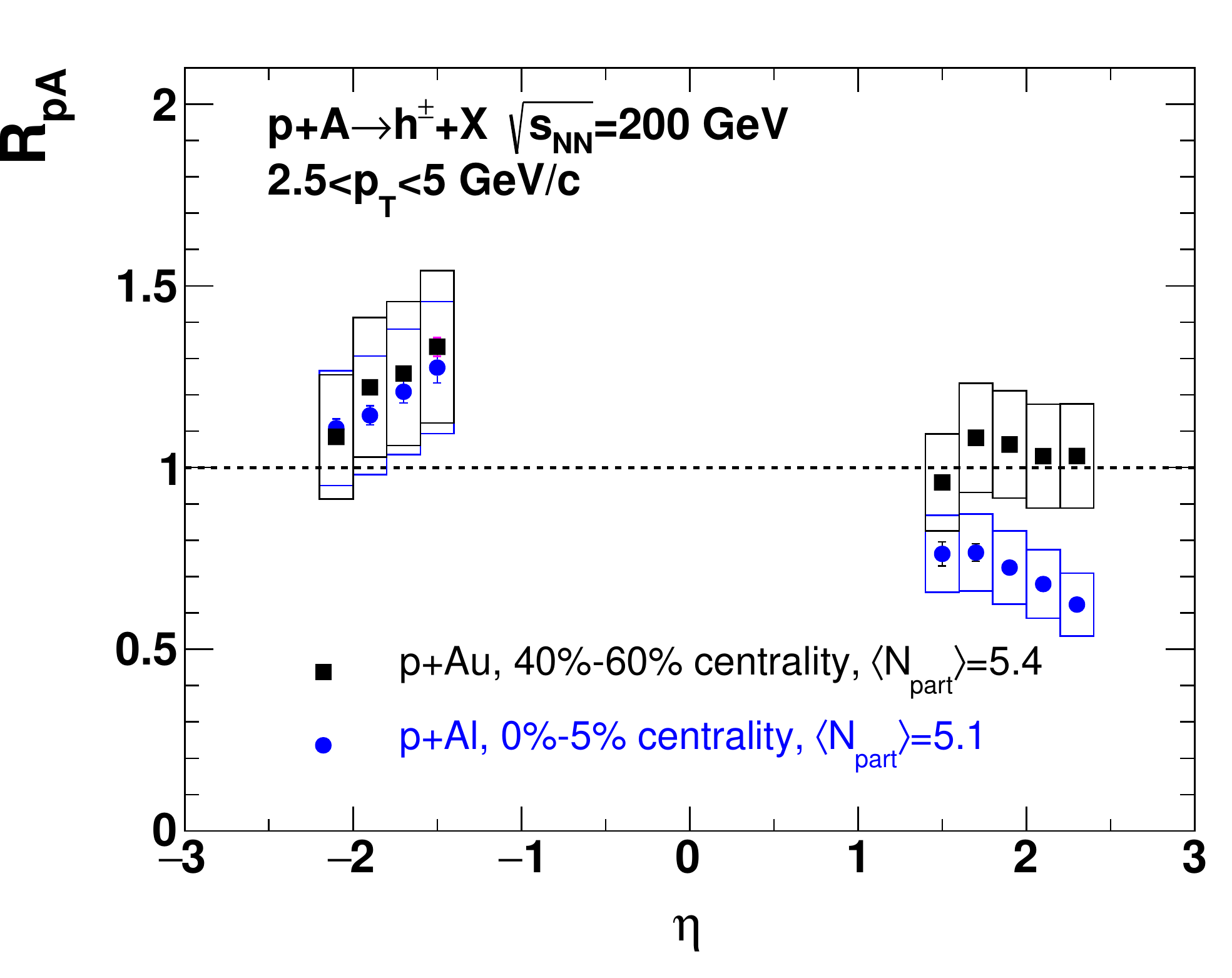}
\caption{\label{fig:rpa_eta_simnpart}
\rpa of charged hadrons in $2.5<p_T<5~{\rm GeV}/c$ as a function of 
$\eta$ in 0\%--5\% \pal and 40\%--60\% \pau collisions at \sqsntwo of 
similar \npart.}
\end{figure}

Figures~\ref{fig:rpa_pT_pal_mb} and~\ref{fig:rpa_pT_pau_mb} show \rpa of 
charged hadrons as a function of \pt at forward and backward rapidity in 
\pal and \pau collisions at \sqsntwo.  Both results in 0\%--100\% 
centrality are obtained by integrating over all centrality and applying 
the bias correction factors.  Bars (boxes) around the data points 
represent statistical (systematic) uncertainties, and boxes around unity 
represent the global systematic uncertainty due to uncertainties in the 
BBC efficiency and the calculated \ncoll. The results for \pal indicate 
that there is little modification at forward rapidity (i.e. in the 
$p$-going direction), whereas a small enhancement is observed in 
$\pt<2~{\rm GeV}/c$ at backward rapidity (i.e. in the Al-going 
direction). In \pau results, a suppression is seen in 
$\pt<3~{\rm GeV}/c$ at forward rapidity unlike the \pal results. At 
backward rapidity, a similar trend of enhancement is observed in the 
\pau data, though with larger magnitude.

Comparisons with estimated \rpa based on nuclear modified PDFs are shown 
from the \ncteq nPDF~\cite{Kovarik:2015cma} and the \epps 
nPDF~\cite{Eskola:2016oht} interfaced with \textsc{pythia 
v8.235}~\cite{Sjostrand:2014zea}; the parameters used in the event 
generation of \pythia are listed in Table~\ref{tab:pythia}. Note that 
the multiplication factor for multiparton interactions is determined by 
comparing the $\eta$-dependent multiplicity distribution in \pp 
collisions at \sqstwo~\cite{Alver:2010ck}. The calculations indicate a 
modest expected suppression at forward rapidity from shadowing of 
low-\xbj partons in the Au nucleus, and are in agreement with the data 
within uncertainties.  However, at backward rapidity, sensitive to 
potential anti-shadowing of higher-\xbj partons in the Au nucleus, the 
calculations result in no modification in contradistinction from the 
data. pQCD calculations considering incoherent multiple scatterings 
inside the nucleus before and after hard scattering~\cite{Kang:2014hha} 
at backward rapidity are also compared with the data, and it agrees with 
the both \pal and \pau data.

Figure~\ref{fig:rpa_eta_mb} shows \rpa of charged hadrons integrated 
over the interval $2.5<\pt<5~{\rm GeV}/c$ as a function of $\eta$ in the 
0\%--100\% centrality selection of (a) \pal and (b) \pau  
collisions at \sqsntwo.  Again the data are compared with pQCD 
calculations at backward rapidity and calculations based on two nPDF 
sets. In \pau collisions, there is a modest hint that enhancement at 
backward rapidity becomes larger as $\eta$ approaches midrapidity, 
while the suppression at forward rapidity becomes stronger. In \pal 
collisions, \rpa at forward rapidity is quite similar to what is 
observed in \pau collisions, whereas it shows a smaller enhancement at 
backward rapidity than the results in \pau collisions. The comparison 
with \ncteq and \epps nPDF calculations indicates that the \rpa at 
forward rapidity agrees in both \pal and \pau collisions, but the 
enhancement at backward rapidity in \pau collisions is not reproduced by 
the both calculations. In case of the comparison with the pQCD 
calculations at backward rapidity, the magnitude of enhancement is 
similar. However, the pQCD calculations show a stronger enhancement at 
more backward rapidity which is different from the trend in the data.

Because initial and final-state nuclear effects on hadron production may 
depend on the density of initial partons in the nucleus and on the 
density of final-state produced particles, \rpa has been measured in 
various centrality bins of \palau collisions. 
Figures~\ref{fig:rpa_pT_cent_pal} and~\ref{fig:rpa_eta_cent_pal} show 
\rpa of charged hadrons as a function of \pt or $\eta$ at forward and 
backward rapidity from the most central bin (0\%--5\%) to the most 
peripheral bin (40\%--72\%) for \pal collisions at \sqsntwo.  The 
results at forward and backward rapidity are plotted together in each 
plot.  First, there is a clear centrality dependence both at forward and 
backward rapidity. The magnitude of the modification, which shows 
enhancement at backward rapidity and suppression at forward rapidity, 
becomes stronger in more central \pal collisions. The observed \rpa in 
the most peripheral (40\%--72\%) \pal collisions is consistent with 
unity in both rapidity regions, indicating little modification of 
charged hadron production compared to the \pp data. Both the magnitude 
of the modification and the \pt dependence are larger in central 
collisions. 
at forward and backward rapidity in central \pau collisions. The 
centrality dependence of \rpa as a function of $\eta$ shown in 
Fig.~\ref{fig:rpa_eta_cent_pal} is consistent with what is seen in \rpa 
as a function of \pt. The $\eta$ dependence at backward rapidity is 
weakly centrality dependent, but there is a clear $\eta$ dependence at 
forward rapidity in the most central collisions.

Figures~\ref{fig:rpa_pT_cent_pau} and~\ref{fig:rpa_eta_cent_pau} show 
\rpa of charged hadrons as a function of \pt and $\eta$ in various 
centrality classes of \pau collisions. Similar to the results in \pal 
collisions, the magnitude of modification becomes larger in more central 
collisions both at forward and backward rapidity, and the \rpa values in 
the most peripheral \pau collisions are consistent with unity. When 
comparing \pal and \pau results in the 0\%--5\% central collisions shown 
in the panel (a) of 
Figs.~\ref{fig:rpa_pT_cent_pal},~\ref{fig:rpa_eta_cent_pal},~\ref{fig:rpa_pT_cent_pau}, 
and~\ref{fig:rpa_eta_cent_pau}, \rpa at forward rapidity is comparable 
between the two collision systems. However, the enhancement at backward 
rapidity is much stronger in \pau collisions.  
Figure~\ref{fig:rpa_pT_cent_pau} compares pQCD calculations with the 
\pau data at backward rapidity. Similarly with the comparison in the 
integrated centrality, the calculation can reproduce the \pt and 
centrality dependent enhancement.

Figure~\ref{fig:rpa_npart} shows \rpa as a function of \npart for 
charged hadrons in the range $2.5<\pt<5~{\rm GeV}/c$ at (a) forward and 
(b) backward rapidity in \palau collisions at \sqsntwo. Unlike the 
previous results, the systematic uncertainty on \ncoll is included in 
boxes around data points. The data show that \rpa at backward rapidity 
(filled [black] circles), i.e. in the $A$-going direction, increases 
monotonically with \npart, and the trend is reproduced by the pQCD 
calculation. However, \rpa at forward rapidity (open [red] circles), 
i.e. in the $p$-going direction, reveal that each collision system has 
its own decreasing trend as \npart becomes larger. \rpa at forward 
rapidity in 0\%--5\% of \pal and \pau collisions are consistent 
($\rpa\sim0.7$), although $\npart$ (9.7 in \pau and 4.1 in \pal 
collisions) are quite different. The trend of a larger enhancement 
(suppression) at backward (forward) rapidity in more central collisions 
is consistent with the previous results of charged hadrons and muons 
from heavy flavor decay in \dau 
collisions~\cite{Adler:2004eh,Adare:2013lkk}. A closer look on 
$\eta$-dependent \rpa in 0\%--5\% \pal and 40\%--60\% \pau collisions of 
similar \npart is shown in Fig.~\ref{fig:rpa_eta_simnpart}. At backward 
rapidity, it shows not only a consistent magnitude of \rpa but also a 
quite similar trend of \rpa in $\eta$. In case of the comparison at 
forward rapidity, \rpa of the 40\%--60\% \pau centrality bin is 
consistent with unity in all $\eta$ bins, whereas a $\eta$-dependent 
suppression is seen in 0\%--5\% \pal collisions.

The suppression of charged hadron production at forward rapidity in 
integrated centrality of \palau collisions can be explained by the nPDF 
modification based on the comparison with the \ncteq and \epps 
calculations shown in 
Figs~\ref{fig:rpa_pT_pal_mb},~\ref{fig:rpa_pT_pau_mb},~and~\ref{fig:rpa_eta_mb}.  
It would be useful to extend another calculation within the CGC 
framework~\cite{Albacete:2010bs}, which successfully describes the 
suppression of charged hadron production at forward rapidity in \dau 
collisions~\cite{Arsene:2004ux,Adams:2006uz}. More differential 
calculations from these various frameworks are needed to compare to the 
systematic trends found in our new results. In addition to these models 
which consider modification of the parton distribution functions inside 
the nucleus, the pQCD calculation of dynamic shadowing considering 
coherent multiple scatterings inside the nucleus~\cite{Qiu:2004da} also 
predicts a rapidity and impact parameter dependent suppression of hadron 
production at forward rapidity. The centrality dependent suppression at 
forward rapidity shown in both \palau collisions also can be described 
by the color fluctuation effects expecting a stronger centrality 
dependence in \pau collisions than \dau 
collisions~\cite{Alvioli:2017wou}. It will be quite useful to have 
theoretical calculations for detailed comparison with the data in \pt, 
rapidity, and centrality.

For the enhancement of charged hadron production observed at backward 
rapidity, estimates from the nPDF sets clearly fail to describe the 
data. A pQCD calculation considering incoherent multiple scatterings 
both before and after hard scattering~\cite{Kang:2014hha}, which can 
describe the enhancement of heavy quark production at backward rapidity 
in \dau collisions~\cite{Adare:2013lkk}, successfully explains the 
centrality and $A$-dependent enhancement. In addition, there is also a 
possibility of hydrodynamic behavior showing a larger elliptic flow of 
charged particles at backward rapidity where the multiplicity is also 
larger than other rapidity ranges~\cite{Adare:2018toe}.

\section{Summary}
\label{sec:summary}

PHENIX has measured the nuclear modification factor \rpa of charged 
hadrons as a function of \pt and $\eta$ at forward and backward rapidity 
in various centrality ranges of \pal and \pau collisions at \sqsntwo. 
The results in central \pal and \pau collisions show a suppression 
(enhancement) in the forward $p$-going (backward, $A$-going) rapidity 
region compared to the binary scaled \pp results of 0.7 (2.0) for \pau 
and 0.9 (1.2) for \pal in $2.5<\pt<5~{\rm GeV}/c$ at a level of 
significance $3.3\sigma$ ($3.2\sigma$) for \pau and $2.7\sigma$ 
($1.1\sigma$) for \pal. In contrast, there is no significant 
modification of charged hadron production observed in peripheral \pal 
and \pau collisions in either rapidity region.
The enhancement 
at backward rapidity shows a clear $A$-dependence, but the suppression 
at forward rapidity is comparable between the two collision systems 
despite more than a factor two larger \npart in \pau collisions. The 
results integrated over centrality are compared to a calculation with 
the \ncteq and \epps nPDF sets. The calculation agrees with the data 
at forward rapidity both in the integrated centrality of \pal and \pau 
collisions, but it fails to describe the enhancement observed at 
backward rapidity in \pau collisions. Because the nPDF sets does not yet 
provide an impact parameter dependent nPDF, the comparison is limited to 
the case of integrated centrality. These data measured in various 
centrality ranges can be useful to test impact parameter dependent nPDFs 
in different nuclei in the future. The pQCD calculation considering 
incoherent multiple scatterings inside the nucleus can describe the data 
at backward rapidity. In addition, a comparison with different models 
can help to improve the understanding of nuclear effects in small 
collision systems.



\begin{acknowledgments}

We thank the staff of the Collider-Accelerator and Physics Departments 
at Brookhaven National Laboratory and the staff of the other PHENIX 
participating institutions for their vital contributions.
We thank Z.-B. Kang and H. Xing for useful discussions and for providing 
theoretical calculations.
We acknowledge support from the Office of Nuclear Physics in the
Office of Science of the Department of Energy,
the National Science Foundation,
Abilene Christian University Research Council,
Research Foundation of SUNY, and
Dean of the College of Arts and Sciences, Vanderbilt University
(U.S.A),
Ministry of Education, Culture, Sports, Science, and Technology
and the Japan Society for the Promotion of Science (Japan),
Conselho Nacional de Desenvolvimento Cient\'{\i}fico e
Tecnol{\'o}gico and Funda\c c{\~a}o de Amparo {\`a} Pesquisa do
Estado de S{\~a}o Paulo (Brazil),
Natural Science Foundation of China (People's Republic of China),
Croatian Science Foundation and
Ministry of Science and Education (Croatia),
Ministry of Education, Youth and Sports (Czech Republic),
Centre National de la Recherche Scientifique, Commissariat
{\`a} l'{\'E}nergie Atomique, and Institut National de Physique
Nucl{\'e}aire et de Physique des Particules (France),
Bundesministerium f\"ur Bildung und Forschung, Deutscher Akademischer
Austausch Dienst, and Alexander von Humboldt Stiftung (Germany),
J. Bolyai Research Scholarship, EFOP, the New National Excellence
Program ({\'U}NKP), NKFIH, and OTKA (Hungary),
Department of Atomic Energy and Department of Science and Technology
(India),
Israel Science Foundation (Israel),
Basic Science Research and SRC(CENuM) Programs through NRF
funded by the Ministry of Education and the Ministry of
Science and ICT (Korea).
Physics Department, Lahore University of Management Sciences (Pakistan),
Ministry of Education and Science, Russian Academy of Sciences,
Federal Agency of Atomic Energy (Russia),
VR and Wallenberg Foundation (Sweden),
the U.S. Civilian Research and Development Foundation for the
Independent States of the Former Soviet Union,
the Hungarian American Enterprise Scholarship Fund,
the US-Hungarian Fulbright Foundation,
and the US-Israel Binational Science Foundation.

\end{acknowledgments}



%
 
\end{document}